\documentclass[12pt,preprint]{aastex}

\usepackage{emulateapj5}
\usepackage{graphicx}
\usepackage{color}
\usepackage{natbib}

\newcommand{\chandra}{{\it Chandra}}
\newcommand{\aox}{\ifmmode \alpha_{ox}\else$\alpha_{ox}$\fi}
\newcommand{\mone}{\ifmmode ^{-1}\else$^{-1}$\fi}
\newcommand{\mtwo}{\ifmmode ^{-2}\else$^{-2}$\fi}
\newcommand{\degs}{\ifmmode ^{\circ}\else$^{\circ}$\fi}
\newcommand{\mv}{\ifmmode {m_{V}}\else${m_{V}}$\fi}

\newcommand{\gae}{\mathrel{\raise .4ex\hbox{\rlap{$>$}\lower 1.2ex\hbox{$\sim$}
} }}

\shorttitle{X-ray emission from GPS/CSS sources}

\shorttitle{X-ray Observations of GPS/CSS Sources}

\shortauthors{Siemiginowska et al.}

\begin{document}

\title{X-ray Properties of the GigaHertz-Peaked 
and Compact Steep Spectrum Sources.}

\author{Aneta Siemiginowska\altaffilmark{1}, Stephanie LaMassa\altaffilmark{1,2}
Thomas L. Aldcroft\altaffilmark{1}, Jill Bechtold\altaffilmark{3}, Martin Elvis\altaffilmark{1}}
\email{asiemiginowska@cfa.harvard.edu}
\altaffiltext{1}{Harvard-Smithsonian Center for Astrophysics}
\altaffiltext{2}{Johns Hopkins University}
\altaffiltext{3}{Steward Observatory, University of Arizona}

\begin{abstract}

We present {\it Chandra} X-ray Observatory observations of Giga-Hertz
Peaked Spectrum (GPS) and Compact Steep Spectrum (CSS) radio
sources. The {\it Chandra} sample contains 13 quasars and 3 galaxies
with measured 2-10~keV X-ray luminosity within $10^{42} -
10^{46}$~erg~s$^{-1}$. We detect all of the sources, five of which are
observed in X-ray for the first time. We study the X-ray spectral
properties of the sample. The measured absorption columns in the
quasars are different than those in the galaxies in the sense that the
quasars show no absorption (with limits $\sim 10^{21} \rm cm^{-2}$)
while the galaxies have large absorption columns ($> 10^{22}
\rm cm^{-2}$) consistent with previous findings. The median photon
index of the sources with high S/N is $\Gamma=1.84 \pm0.24$ and it is
larger than the typical index of radio loud quasars. The arcsec
resolution of {\it Chandra} telescope allows us to investigate X-ray
extended emission, and look for diffuse components and X-ray jets. We
found X-ray jets in two quasars (PKS~1127-145, B2~0738+32), an X-ray
cluster surrounding a CSS quasar (z=1.1, 3C~186), detected a possible
binary structure in 0941-080 galaxy and an extended diffuse emission
in galaxy PKS~B2~1345+12. We discuss our results in the context of
X-ray emission processes and radio source evolution. We conclude that
the X-ray emission in these sources is most likely unrelated to a
relativistic jet, while the sources' radio-loudness may suggest a high
radiative efficiency of the jet power in these sources.

\end{abstract}

\section{Introduction}

Powerful radio sources exemplify the most energetic processes in the
Universe and demonstrate that an accreting supermassive black hole can
influence regions at megaparsec distances away from its immediate
sphere of influence. How these sources are triggered and how long they
last are subjects of many scientific investigation. In this paper we
present X-ray observations of a sample of the most powerful, but
compact radio sources. They may represent an early stage of just
triggered quasar activity. Giga-Hertz Peaked Spectrum (GPS) and
Compact Steep Spectrum (CSS) radio sources typically have compact
radio morphology with the radio emission contained within their host
galaxy ($< 10$~kpc, O'Dea 1998).  The compact radio structure on
miliarcsec scales is similar to the morphology of a large radio source
with lobes, hot spots and jets. Because of this similarity and the
observed high radio power the GPS sources are thought to be the
precursors of large radio galaxies observed at the early stage of
their expansion ({\it evolution model}, Fanti et al 1995, O'Dea \&
Baum 1997, O'Dea 1998). Note that an alternative model, where the
source is confined by a dense environment ({\it frustrated}) has not
been completely ruled out \citep{snellen,alex2000}, although there is
no evidence for the medium with the required column density
\citep{morganti2007}

If the evolutionary hypothesis is correct the studies of GPS/CSS may
impact our understanding of triggering quasars. Measured expansion
velocities of double radio structures \citep{polatidis2003} in several
nearby GPS galaxies support the idea of them being young
($<10^4$~years). In addition their radio morphology suggests that the
sources are observed at high inclination angles, so the Doppler
beaming is not important.  Thus the young age of {\it GPS galaxies}
with the double or symmetric radio morphology has not been questioned
and evolution studies often consider only the GPS/CSS galaxies,
e.g. the sources where the central AGN engine cannot be directly
observed as it is might be buried by a large column of obscuring
material \citep{guainazzi2006}.

Luminous quasars are thought to be powered by high accretion rates and
radio loud quasars in addition to strong thermal emission also exhibit
relativistic jets carrying large kinetic power.  {\it GPS/CSS quasars}
with a core-jet radio morphology provide a challenge for measuring
their age because a truly young GPS/CSS quasar could often be confused
with a blazar observed along the jet axes
\citep{lister, stan2005}.  However, there should also be young sources
among blazars.  Identifying typical blazars and young quasars within
the GPS/CSS quasar samples is not trivial and has been basically
avoided.  In order to distinguish a blazar from a young radio source
one needs to consider the entire broad-band spectrum of each
source. This is hard and requires simultaneous observations at many
frequencies, as blazars typically vary. A correlated rapid variability
across many frequencies including the gamma-ray band may indicate a
blazar nature, while a young GPS/CSS source should not vary on the
short timescales (less than a few months).

X-ray emission is predicted as a result of the radio source evolution
and its expansion into the surrounding interstellar and intergalactic
medium (ISM/IGM) \citep{heinz1998}.  A detection of such emission
could provide information about temperature and density, e.g. the
physical conditions of the expanding radio source.

While there has been an abundance of the radio data collected over the
last decade, X-ray observations of GPS/CSS sources have been sporadic.
O'Dea (1998) lists 31 GPS sources ((only quasars were detected while 7
galaxies had upper limits) showing quite high X-ray luminosities
(L$_{X(0.2-4.5\,keV)}
\sim 10^{45}-10^{46}$ergs~s$^{-1}$).  An intrinsic X-ray absorption
was reported by \citet{elvis94}  in two out of three high redshift
GPS quasars suggesting that their environment might be different from
that of the other quasars.  ROSAT upper limits for a few GPS/CSS
galaxies of L$_X < 3 \times 10^{42}$ ergs~s$^{-1}$ are consistent with
the X-ray emission expected from poor clusters or early type galaxies.
The first X-ray detection of a GPS galaxy at L$_X \sim 2
\times 10^{42}$ ergs~s$^{-1}$ by ASCA was reported by O'Dea et al (2000).
Recent XMM-{\it Newton} observations of a few GPS galaxies
(\cite{guainazzi2006}, \cite{vink2006}) indicate an intrinsic
absorption with an average column of $\sim 10^{22}$~cm$^{-2}$, and the
radio-to-X-ray luminosity ratios comparable to the normal radio
galaxies. There is also no compelling evidence for a hot gas in the
X-ray spectra of GPS galaxies. These observations support the
evolution model in which GPS galaxies are ''young'' counterparts for
large-scale FRII galaxies \citep{guainazzi2006}.

The highest spatial resolution X-ray observations before the launch of
the {\it Chandra} X-ray Observatory ({\it Chandra}) were made with the
ROSAT HRI in which 2 out of 4 GPS quasars showed traces of an extended
emission (Antonelli \& Fiore 1997, Siemiginowska et al 2003).  The
spatial resolution of {\it Chandra} enables to detect X-ray structures
on arcsec scales with a high dynamic range \citep{weisskopf2003}.
During the first years of the mission we have made a surprising
discovery - the two of the GPS quasars have shown hundred kilparsec
scale X-ray jets in the {\it Chandra} X-ray images (\cite{siem2002},
\cite{siem2003}). These two quasars have been classified as GPS
quasars in radio band with no indication of a large scale
emission. However, a careful re-analysis of archival radio data show
that there was indeed a large scale radio structure present, but due
to a strong quasar core emission only detectable in the high dynamic
range radio data.

For this exploratory X-ray study we selected 16 GPS/CSS sources from
the sample presented in \citet{odea1998}. Our X-ray sample is not
complete by any means, but it is so far the largest sample of GPS/CSS
sources with the highest quality X-ray data.  The main goal of this
study is to obtain the X-ray flux and spectra, and to investigate an
extended X-ray emission, disentangle diffuse and nuclear components,
look for X-ray jets and study absorption properties.  We found X-ray
jets in two quasars, discovered an X-ray cluster surrounding a
redshift z=1 CSS quasar, detected a possible binary structure in one
galaxy and an extended diffuse emission in the other.  Overall the
absorption properties of GPS quasars are different than those of GPS
galaxies in the sense that the quasars show no absorption (with limits
$\sim 10^{21} \rm cm^{-2}$) while the galaxies have large absorption
columns consistent with the previous findings.

The paper is structured in a following way. We present the X-ray
sample in Section 2; in Section 3 we show the X-ray Chandra data and
describe the data analysis process; in Section 4 we present the
results and Section 5 contains the discussion of the results. We
summarize the results in Section 6.  Throughout this paper we use the
cosmological parameters based on the WMAP measurements (Spergel et
al. 2003): H$_0=$71~km~sec$^{-1}$~Mpc$^{-1}$, $\Omega_M = 0.27$, and
$\Omega_{\rm vac} = 0.73$.

\section{X-ray Sample}

Our original {\it Chandra} sample of GPS/CSS sources was chosen from
the complete sample compiled by \cite{odea1998} that consists of two
samples studied by \cite{fanti1990} and \cite{stan1998}. Fanti et al
sample was derived from 3C catalog with the following criteria: (1)
flux density larger than $>10$Jy at 178~MHz; (2) projected linear size
$> 20$~kpc; (3) $|b|> 10$deg and $|l|> 10$deg; (4) log power at
178~MHz$> 26.75$~W~Hz$^{-1}$.  The Stanghellini et al. (1998) sample
was selected from the complete catalog of radio sources stronger
than 1~Jy at 5 GHz compiled by Khur et al.(1981) with the following
criteria: (1) a source flux density measured at 5~GHz should be larger
than $>$1~Jy; (2) a radio spectrum has a turn-over at the frequency
between 0.4-6~GHz; (3) a spectral index is steep, $\alpha >$0.5 at
high frequency end, after the turn-over; (4) a source declination is
above $ > -25\deg$ and galactic latitude above $> 10 \deg$.  The radio
spectra for this sample were derived from simultaneous observations to
prevent ambiguity from variability. Also, no regard was given to the
optical counterpart -- quasars and galaxies are both included.  This
means that the sample can be used to investigate the GPS phenomenon in
a range of galaxy environments.

The 11 GPS/CSS sources were selected  based on the existing X-ray
detections, the size of the radio emission and the low Galactic
equivalent Hydrogen column. 
Four other sources from  \cite{odea1998} sample have been already observed by
\chandra\, and we obtained the data from the archive. We also added one more
GPS source available in the archive, although it has not been on the
\cite{odea1998} list.  
The main purpose of this study is to investigate the X-ray properties
of the GPS/CSS sources and we decided to expand the samples by the
archival data.

\begin{figure}
\includegraphics[width=8cm, angle=90]{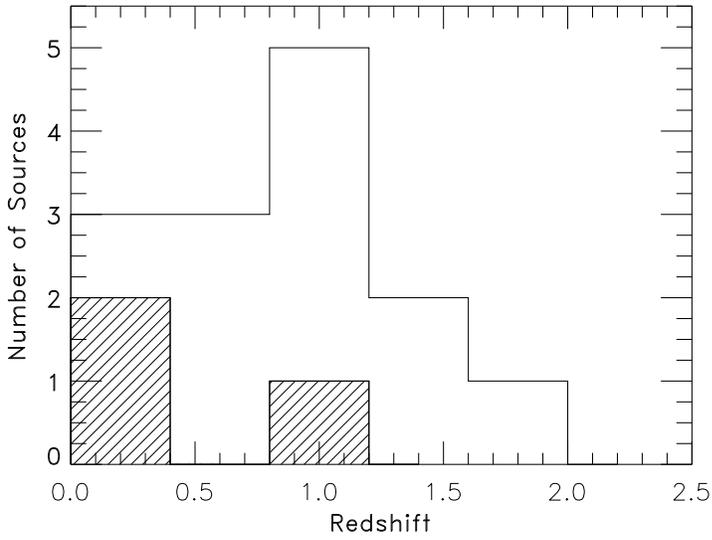}
\caption{Histogram of number of sources vs. redshift.  Galaxies are represented
by filled region.}
\label{redshift}
\end{figure}

The final \chandra\, sample presented here contains 3 GPS galaxies and
13 GPS/CSS quasars. The sample is listed in Table~\ref{tab1} with the
date of each observation, observation IDs (obsid) and the exposure
times.  Eleven sources have been previously observed in X-rays. The
\chandra, exposure time was estimated assuming a detection of the
extended emission 10 times fainter than the core emission.  Short
$\sim$5~ksec exposures were obtained for the 5 sources that have never
been observed in X-rays to determine their count rates and X-ray
luminosity. The sample spans moderate redshifts (from z=0.228 to
z=1.95, Fig.~\ref{redshift}) and radio luminosities between
L(5GHz)$\sim 10^{43}-10^{45}$ergs~s$^{-1}$ with one GPS galaxy below
L(5GHz)$\sim10^{42}$ergs~s$^{-1}$.

\section{X-ray Observations and Data Analysis}

The {\it Chandra} ACIS-S data were collected in two different ways:
(1) long exposures ($>10$~ksec) to allow detection of a diffuse X-ray
emission on arcsec scales; (2) short observations ($\sim 5$ ksec) to
detect X-ray emission and obtain an X-ray flux for sources with no
previous X-ray detections. All observations were performed with the
source located 30~arcsec from the default aim-point position on the
ACIS-S backside illuminated chip S3 (Proposers' Observatory Guide,
POG\footnote{http://cxc.harvard.edu/proposer/POG/index.html}). Most
data were collected with the 1/8 subarray readout mode of only one CCD
to mitigate pileup, however some of the archival data were taken in
the full readout mode and are affected by up to $\sim 20\%$ pileup.

All 16 sources were detected by {\it Chandra} with number of 
counts between 9 and 14,800.

The X-ray data analysis were performed in CIAO 3.4 with the
calibration files from CALDB 3.3 data base. Note that these
calibration files account for ACIS-S contamination.  Although the
pileup fractions are relatively low we include the pileup model
specified by Davis (2000) and implemented in {\it Sherpa} \citep{sherpa}
in our analysis to recover the intrinsic source continuum for
a few sources the most affected by pileup (see details below).

\subsection{Imaging Analysis}

The ACIS-S images were inspected for the presence of an extended X-ray
emission. The data show extended components in the vicinity of the
X-ray core for 2 sources: a quasar 0740+380 and a
galaxy B2~1345+125. We used the CHART simulator to obtain the PSF for
these sources and confirmed the presence of the extended component.  A
detailed analysis and a discussion of the properties of the X-ray
cluster emission with a total of 740 counts detected out to
$\sim$120~kpc from the quasar 0740+380 (z=1.063) have been presented in
Siemiginowska et al (2005).

\begin{figure*}
\includegraphics[width=8cm]{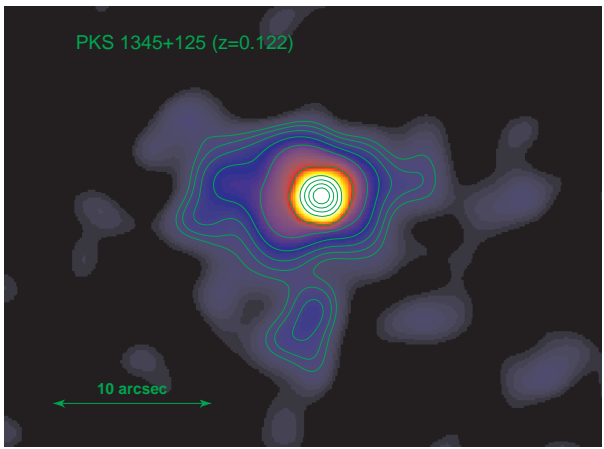}
\caption{ACIS-S image of PKS~B2 1345+125 smoothed using a Gaussian kernel. 
The scale is marked in the lower left corner. North is up and East to the left.}
\label{1345}
\end{figure*}

Fig.~\ref{1345} shows a smoothed ACIS-S image of the galaxy,
PKS~B2~1345+125 (z=0.122) with the extended emission on the $\sim
10$~arcsec scale ($\sim 20$~kpc). A detailed analysis of this
structure is given in Guainazzi et al (in preparation).

\subsection{Spectral Analysis}

We have used CIAO 3.4 and CALDB 3.3 to analyze all data sets, using
the CIAO default tools to extract the spectrum and the associated
calibration files ({\tt rmf} and {\tt arf}). We used
Yaxx\footnote{http://cxc.harvard.edu/contrib/yaxx/package} to uniformly
analyzed and fit the data. The spectra were extracted from circular
regions centered on each source. Annuli surrounding the source regions
were assumed for the background regions.  The total counts detected
for each source are listed in Table~\ref{tab1}.

We fit an absorbed power law model to the spectral data:

$N(E)= A E^{-\Gamma}*\rm exp^{- N^{gal}_H \sigma (E) - N^{z_{abs}}_H
\sigma(E(1+z_{abs}))}$
~photons~cm$^{-2}$~sec$^{-1}$~keV$^{-1}$, 
\noindent
where $A$ is the normalization at 1~keV, $\Gamma$ is the photon index
of the power law and $N^{gal}_H$ and $N^{z_{abs}}_H$ are the two
components for the absorption. The first absorption component is the
effective Galactic absorption characterized by the equivalent neutral
Hydrogen column N$^{gal}$ which we list for each source in
Table~\ref{tab1}
(COLDEN\footnote{http://cxc.harvard.edu/toolkit/colden.jsp}). This
absorption was constant during fitting. We assume that the second
absorption component is intrinsic to the quasar and located at
redshift $z_{abs}$, with N$^{z_{abs}}_H$ as the equivalent hydrogen
column. $\sigma (E)$ and $\sigma (E(1+z_{abs}))$ are the corresponding
absorption cross sections (Morrison \& McCammon 1983, Wilms, Allen and
McCray 2000). We used Powell optimization
and $\chi^2$ (data variance) statistics (binning the data to contain
a minimum of 16 counts in a bin) to determine the best fit parameter
values.  
The modeling results are presented in the Table~\ref{tab2}.

The above model was applied to each source to obtain a uniform
description of the sample. For sources with a high count rate and high
number of counts we also included the effects of pileup in our
modeling.

\subsubsection{Pileup Analysis}

The pileup fractions for the observations were estimated 
a plot of pileup fraction versus photon/frame in Section~6.14 of the
POG$^2$. The photon/frame parameter was calculated by dividing the
number of counts by the exposure time and multiplying this ratio by
the frame time for each observation.  These estimates indicated that
pileup was significant in 
Q0134+329, and Q1416+067.

In order to correctly model pile-up, we removed the ACIS afterglow
correction applied in the standard data processing (SDP).  This
correction, intended to discard cosmic ray events, could also remove as
much as 20\% of valid source photons.  

In \textit{Sherpa}, an absorption (\textit{xsphabs}) and power law
(\textit{pow}) model were fit to the data within the 0.3 to 7.0 keV
energy range; Q1416+067 also included a redshifted absorption
(\textit{xszwabs}) component since N$_H$, the absorption column was
detected in the previous fit for this source.  A pileup model
(\textit{jdpileup}) was included in these fits, using the Monte-Powell
fitting method.

\begin{figure}
\includegraphics[width=8cm, angle=90]{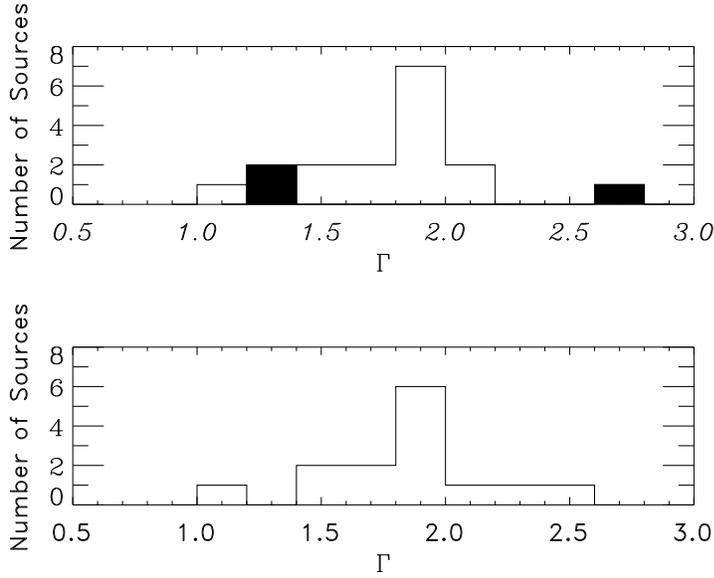}
\caption{Histogram of number of sources vs. photon index $\Gamma$. 
Top plot represents all sources in the sample, with the 3 galaxies
denoted by the filled regions. The bottom plot contains only quasars,
and {\tt jdpileup} pileup model was applied to the 3 sources with the
significant pile-up (see Table 2).}
\label{gamma}
\end{figure}

The results of these fits are shown in Table~\ref{tab1} which displays the
spectral index $\Gamma$ obtained from adding in the pileup component
as well as the estimated pileup fraction and the pileup fraction found
from running the fit.  For each target, the index $\Gamma$ increased
after including the pile-up component in the fit as expected.  The
pileup fraction calculated through {\it Sherpa} was significantly
higher than that estimated from the POG for 
Q0134+329. For the quasars in the sample, the average photon index
increased to 1.84${\pm 0.06}$ from 1.77${\pm 0.06}$ when the pileup
model was included (see Fig.~\ref{gamma}).

Due to the strong pileup (it modifies the PSF) it is hard to determine
whether any extended X-ray emission component is present in these
sources.

\section{Results}

In general the GPS/CSS sources are strong X-ray point sources, however
an extended X-ray emission is detected in several cases in a form of
large scale jets or a diffuse component. We discuss this aspect of our
studies in the next section in more detail.

\begin{figure}
\includegraphics[width=1.8in]{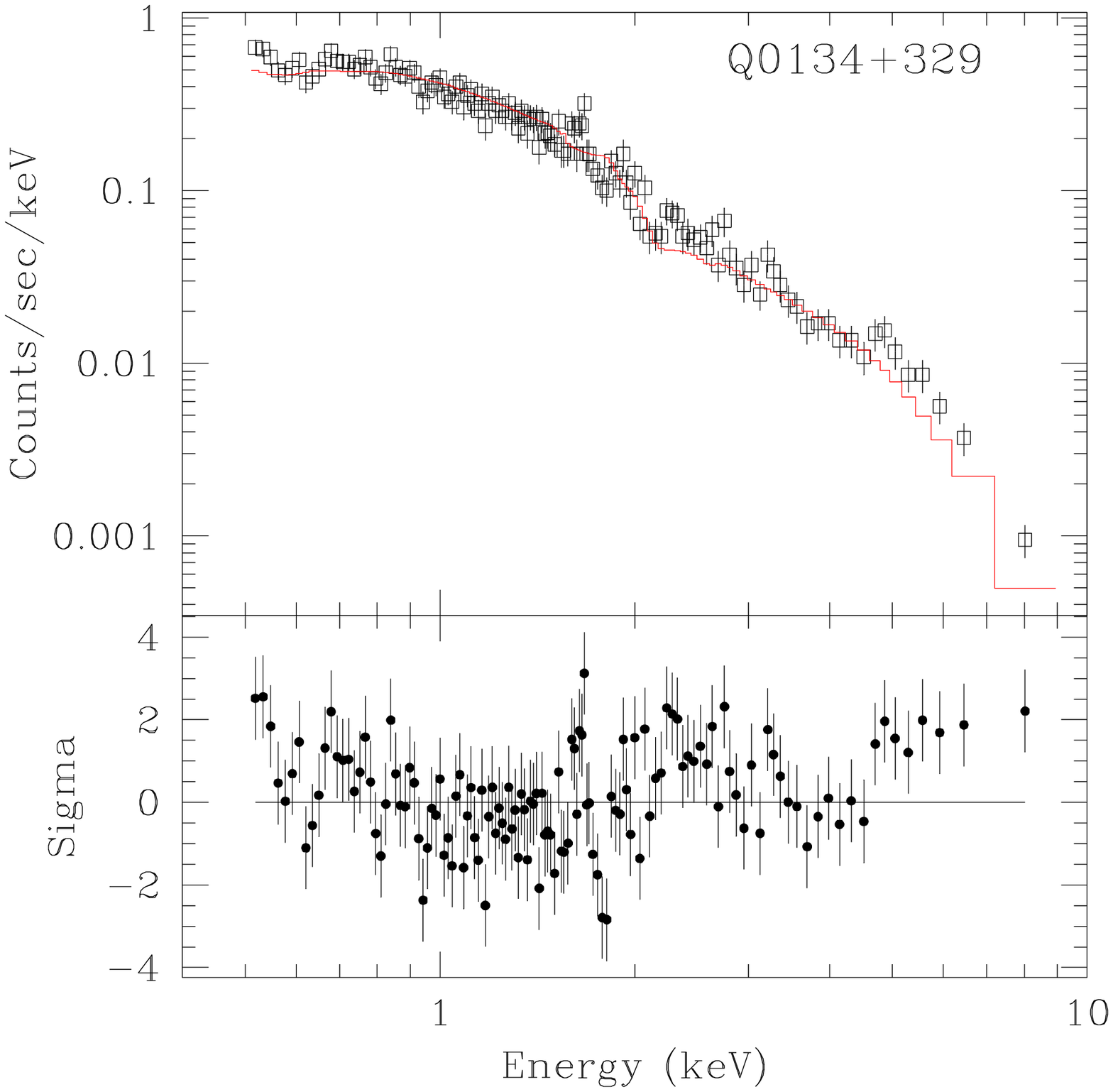}
\hfill
\includegraphics[width=1.8in]{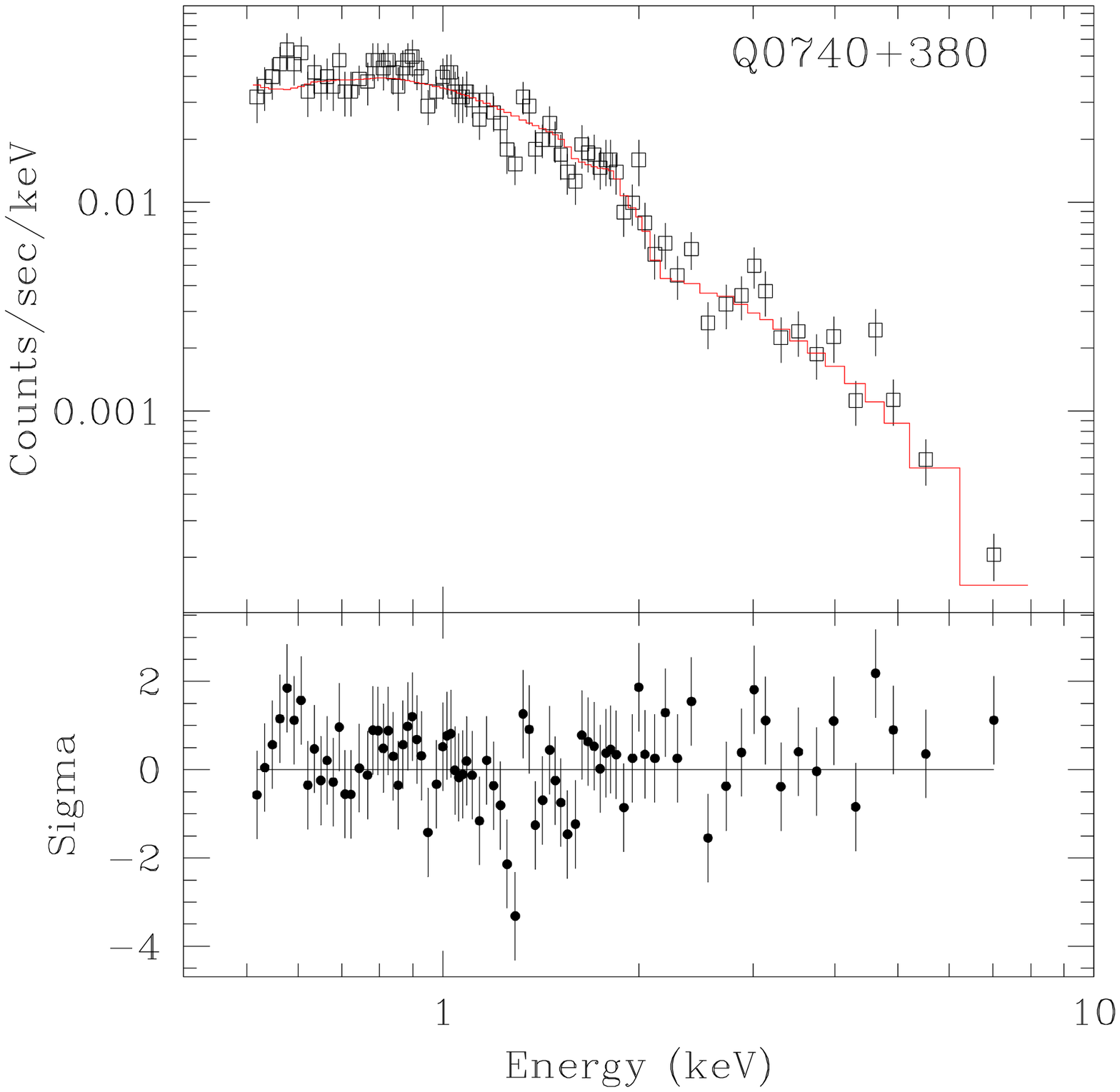}
\hfill
\includegraphics[width=1.8in]{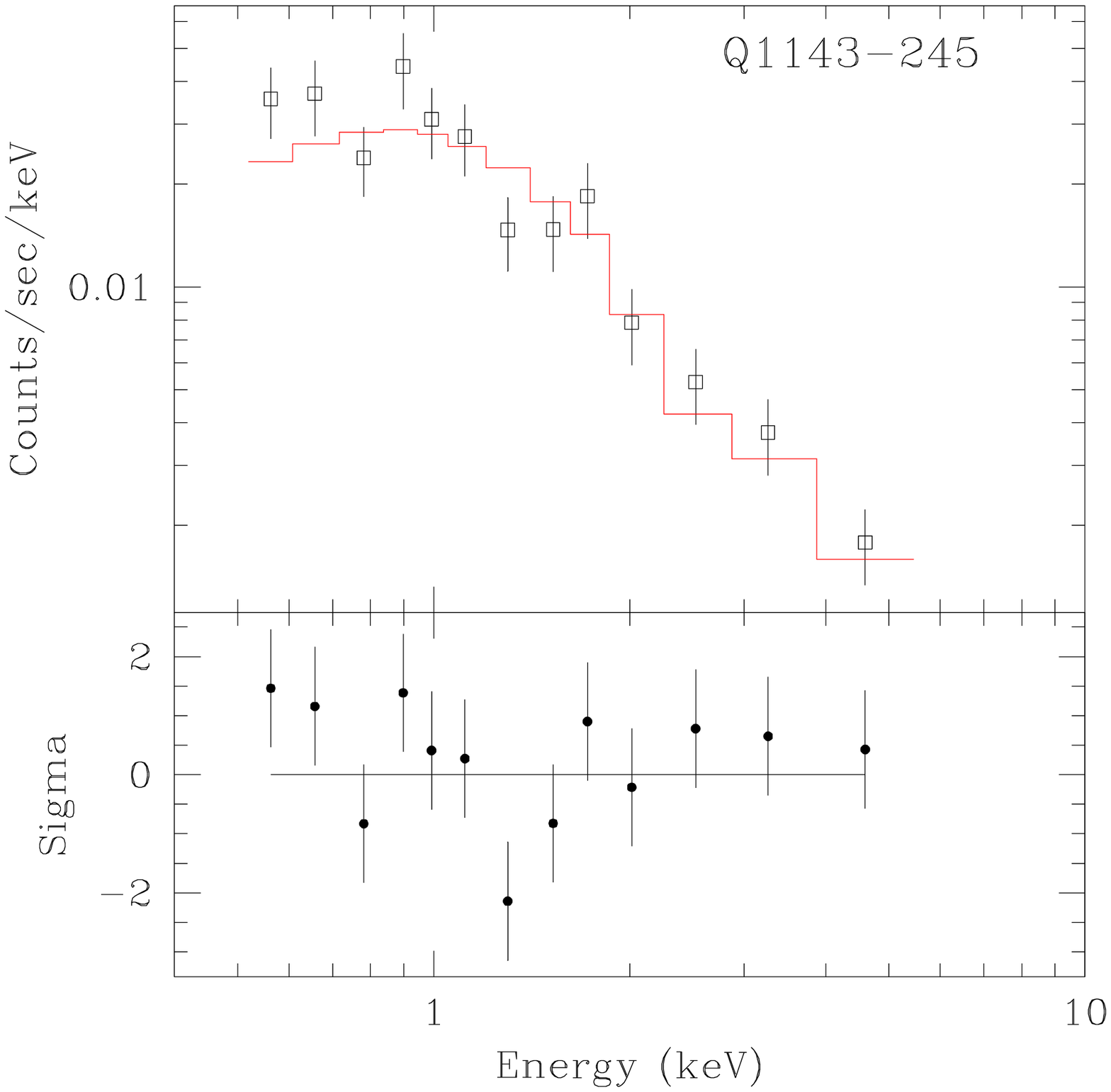}
\hfill
\includegraphics[width=1.8in]{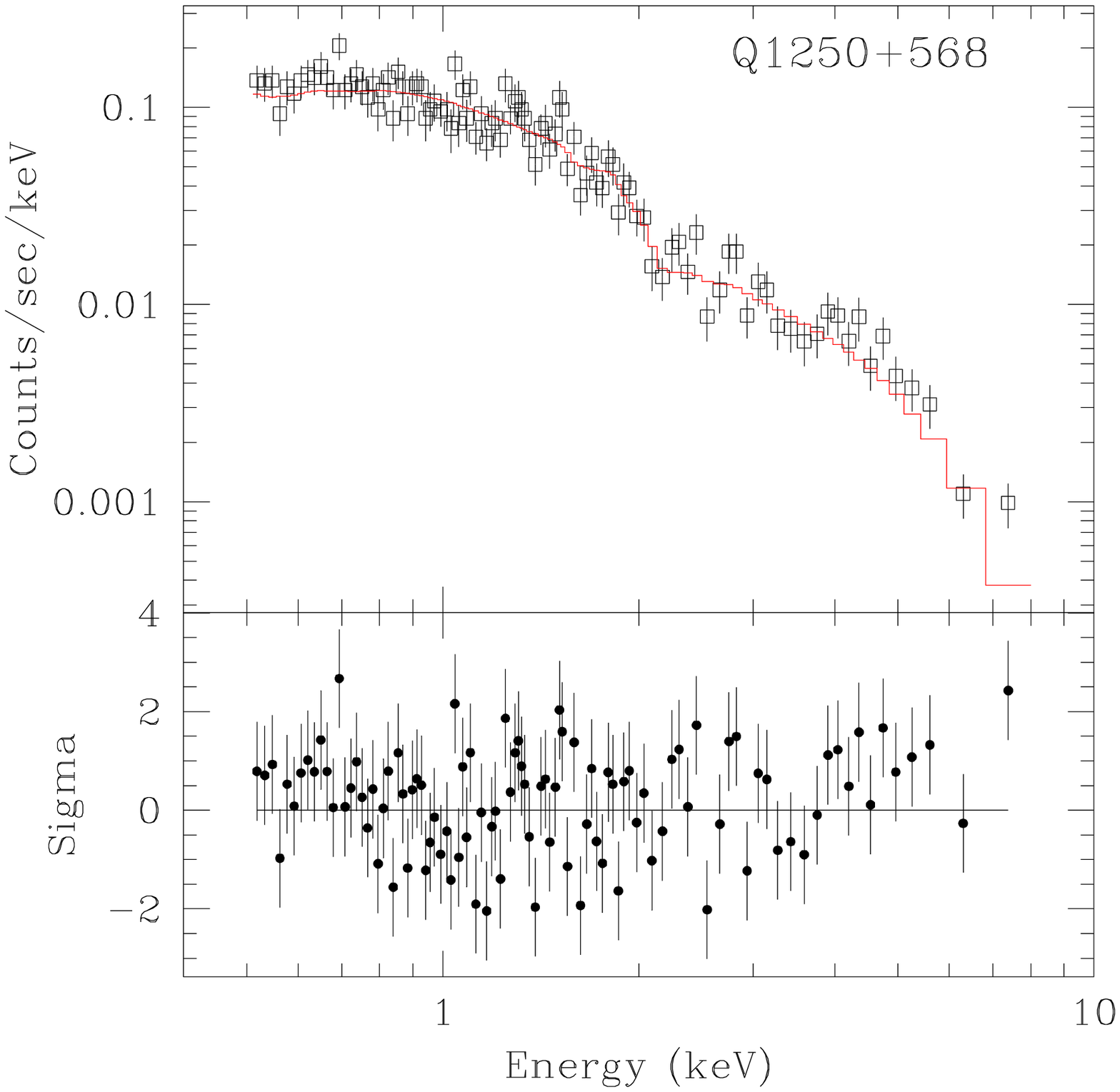}
\hfill
\includegraphics[width=1.8in]{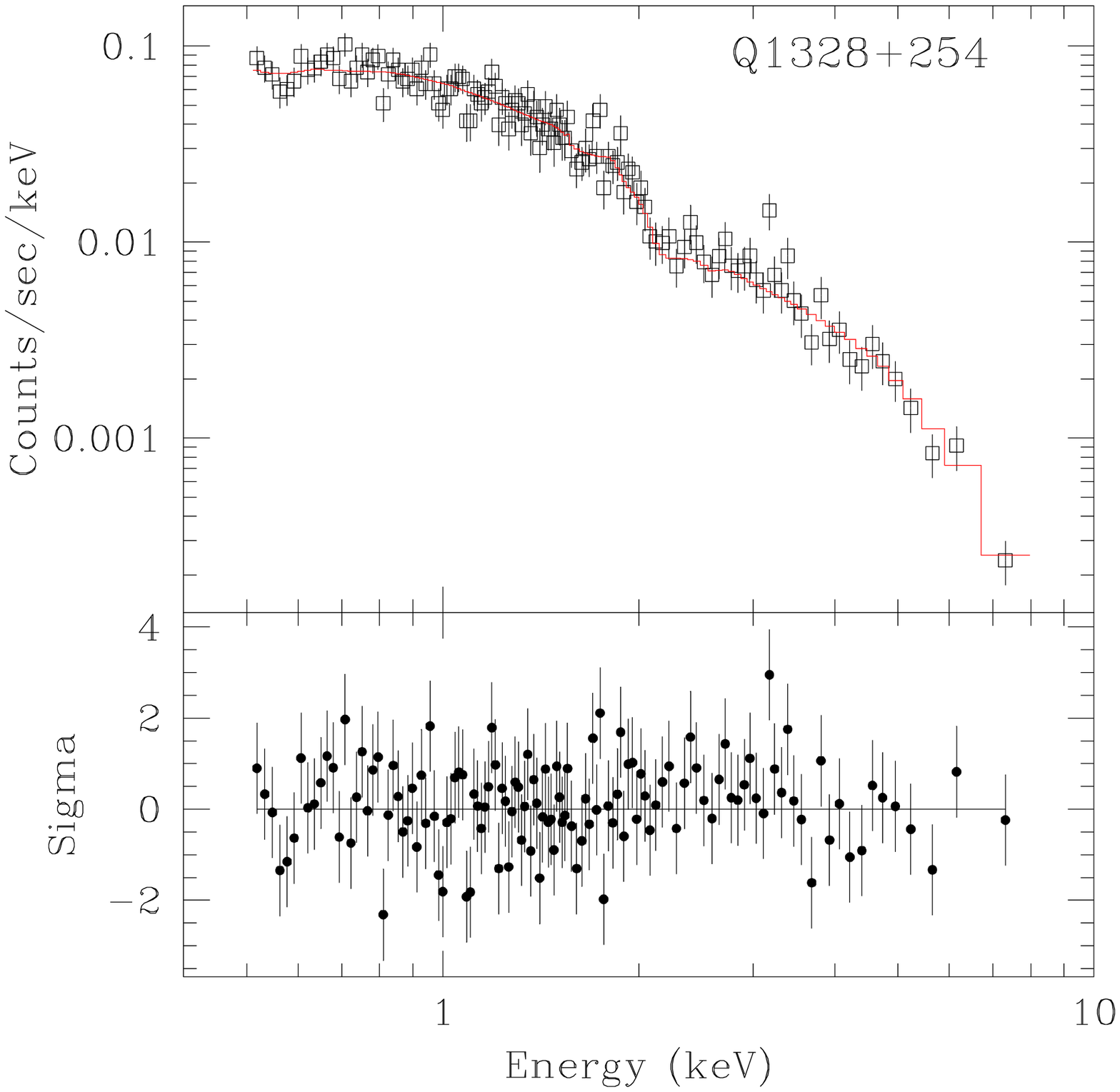}
\hfill
\includegraphics[width=1.8in]{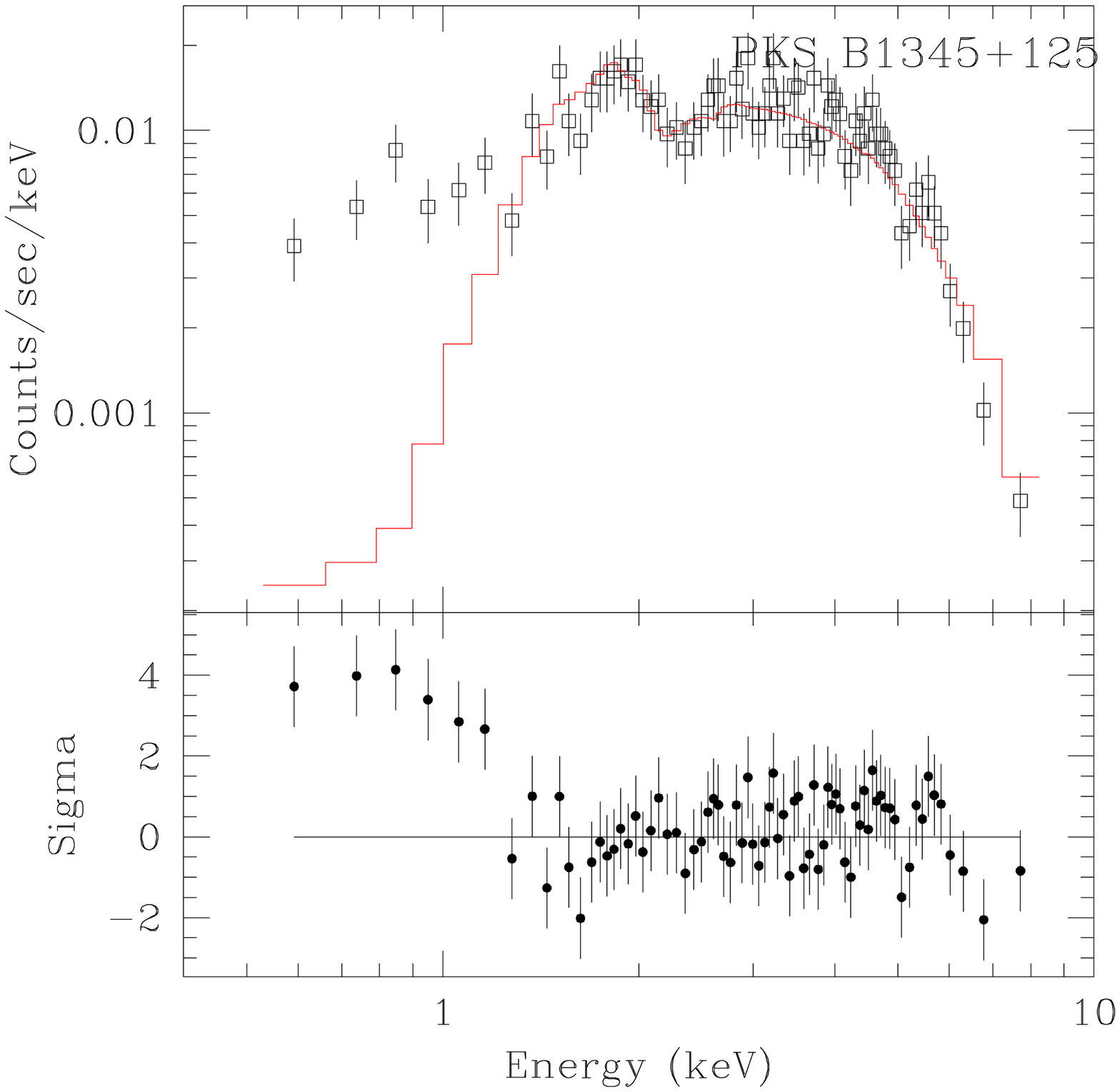}
\hfill
\includegraphics[width=1.8in]{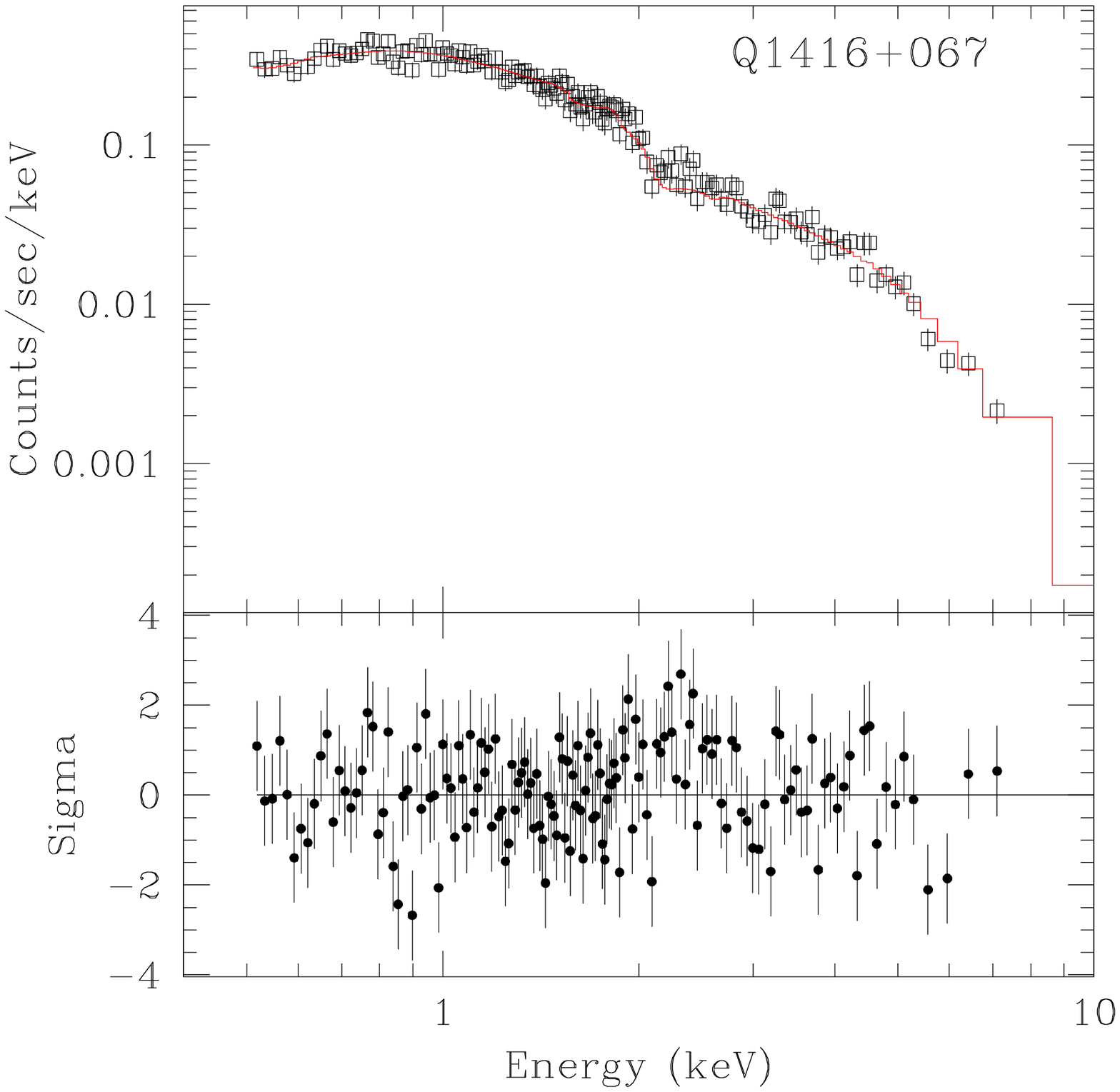}
\hfill
\includegraphics[width=1.8in]{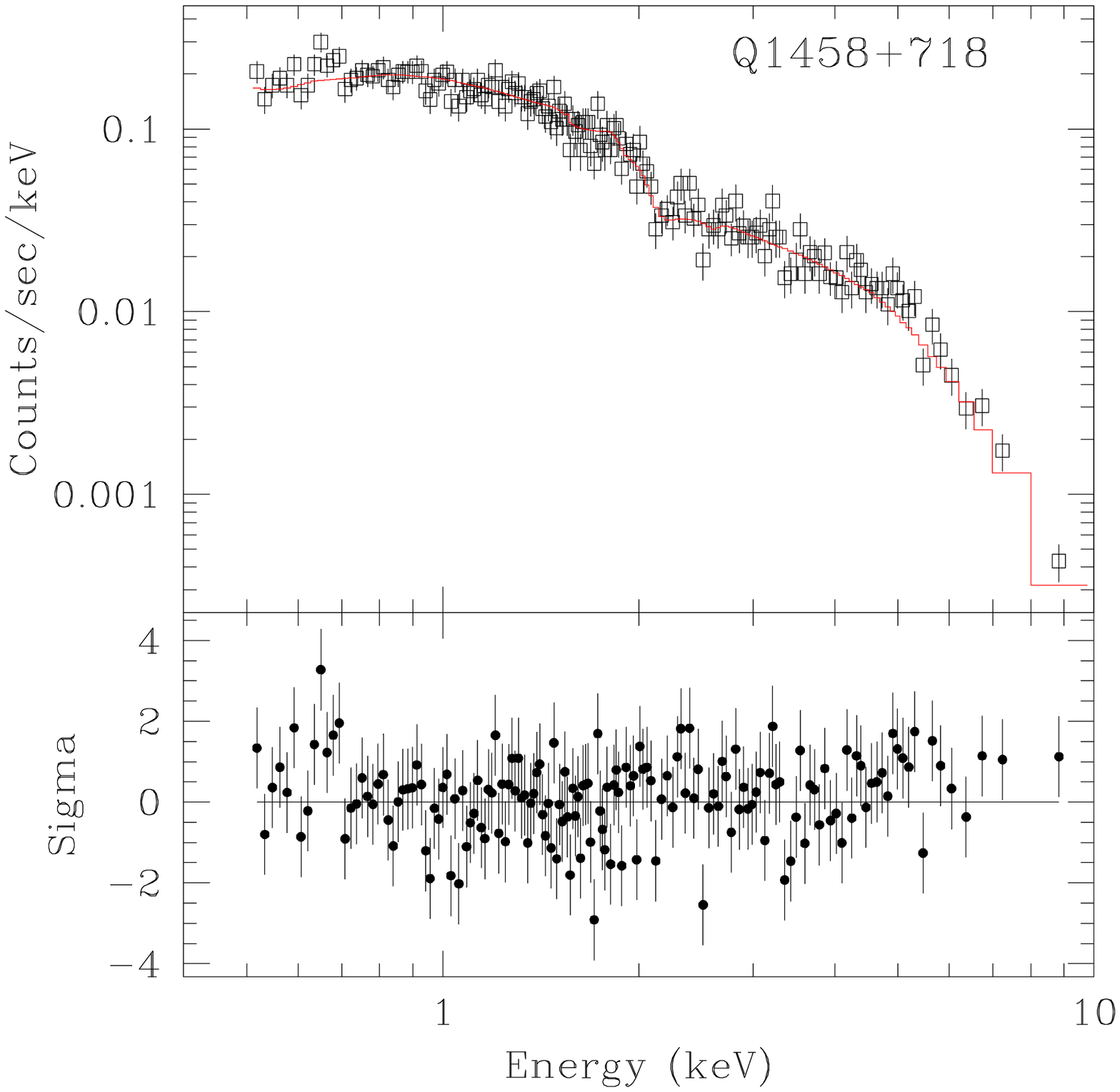}
\hfill
\includegraphics[width=1.8in]{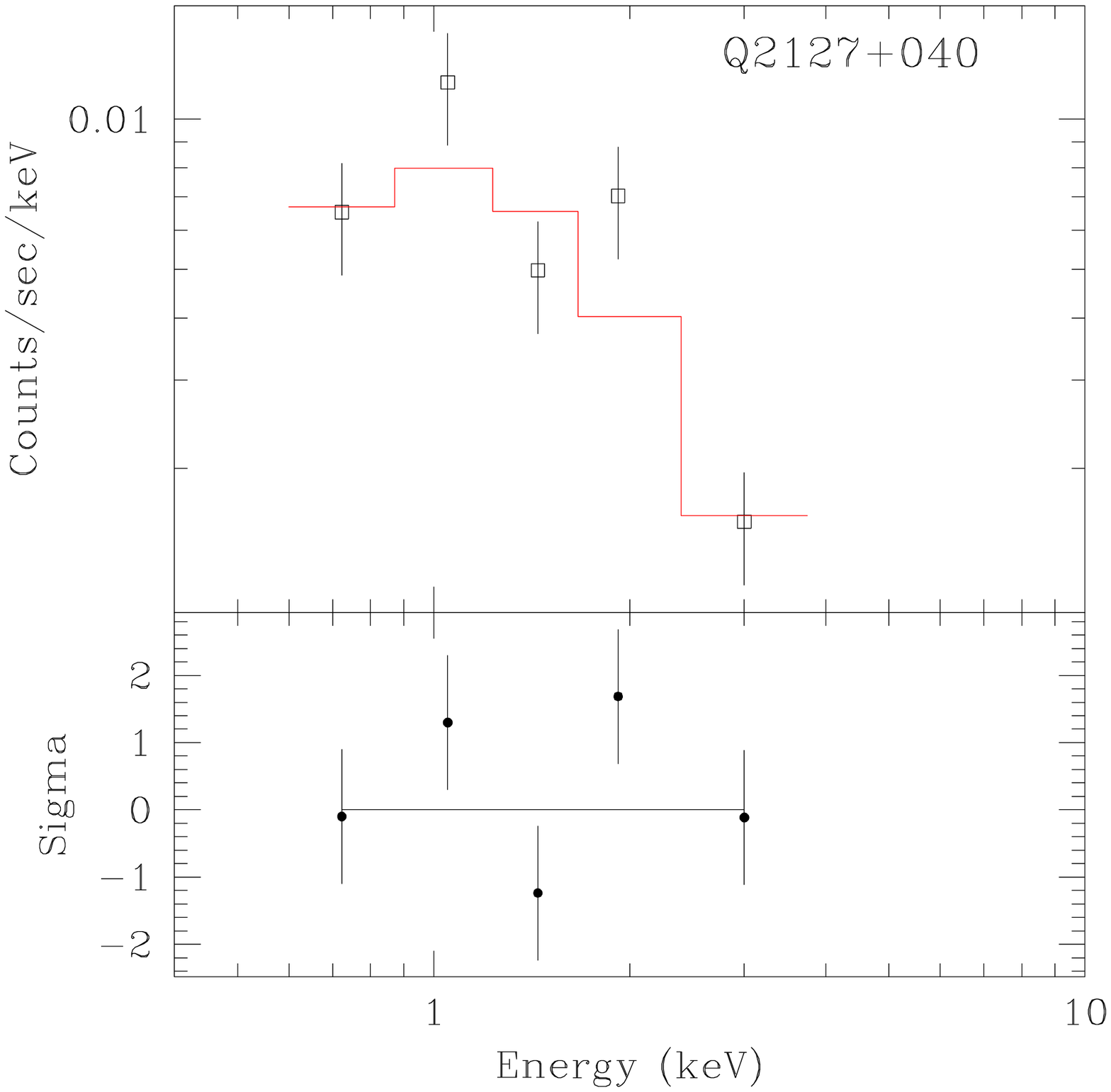}
\hfill
\includegraphics[width=1.8in]{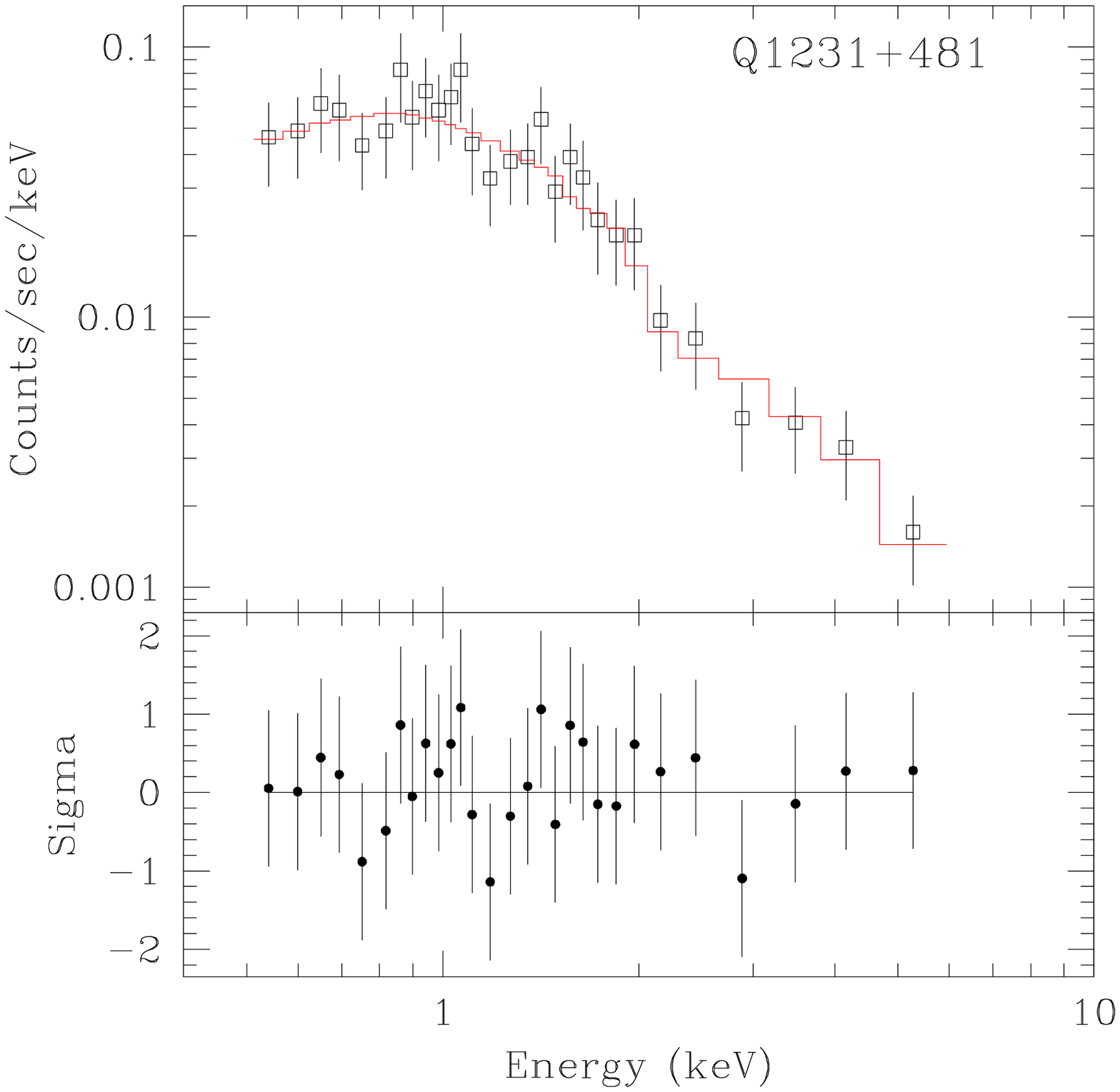}
\caption{\chandra, ACIS-S spectra of GPS/CSS sources fit with the absorbed power law model. The lower panel shows the residuals in terms of sigma. 
 From top left: 0134+329, 0740+380, 0941-080, 1143-245, 1245-197,
 1250+568, 1328+254, B2~1345+125, 1416+067, 1458+718, 1829+29,
 2128+048.}
\label{spectra}
\end{figure}

Fig.~\ref{spectra} plots the absorbed power law model fits to the
X-ray spectra for each source with the residuals.
Table~\ref{tab2} shows best fit values of the absorption column and
photon index with 90\% errors for one significant parameters. All
sources are relatively well described by the power law model.
However, the $\chi^2$ values reflect the scatter observed in the
residuals in Fig.~\ref{spectra} and indicate the need for a more
complex spectrum than a simple power law model for a few sources. For
example PKS~B2~1345+125 has a large absorption column (N$_H
=2.54^{+0.63}_{-0.58} \times 10^{22}$~cm$^{-2}$) and a soft X-ray
excess, Q0134+329 is better described by a two component power law
model, while Q0740+380 shows a possible absorption line (but in this
case the signal to noise is low in this part of the spectrum and
calibration uncertainties are large enough that we cannot claim this
line detection).

We applied a two component model to the PKS~B2~1345+125 X-ray
spectrum, e.g.  an absorbed power law and a bremstrahlung emission, to
determine the luminosity of the soft X-ray thermal emission. The
parameters of this model are the following: the column density of the
intrinsic absorber equal to N$_H =2.28^{+0.36}_{-0.35} \times
10^{22}$~cm$^{-2}$, photon index of the absorbed power law $\Gamma =
1.27^{+0.21}_{-0.19}$ (slightly different than in the previous single
power law fit), the temperature of the thermal component
kT$=0.05^{+0.02}_{-0.02}$~keV and the flux within 0.5-2~keV of 5.3$\times
10^{-14}$~ergs~cm$^{-2}$~s$^{-1}$ resulting in the observed luminosity
of 9.5$\times 10^{42}$~ergs~s$^{-1}$.  The unabsorbed flux in
0.5-2~keV energy range from the power law component results in
$\sim 10^{46}$~ergs~s$^{-1}$ luminosity which is comparable to a
typical quasar luminosity.

\begin{figure}
\includegraphics[width=8cm, angle=90]{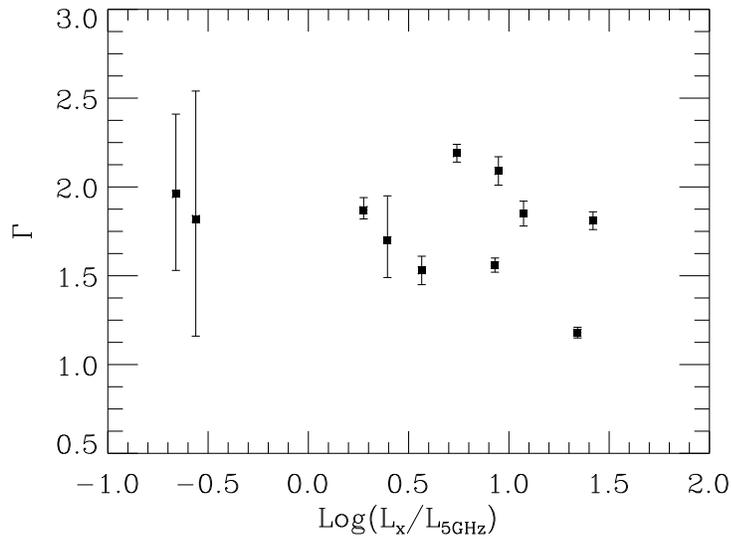}
\caption{Photon index $\Gamma$ vs.  X-ray to radio luminosity ratio for the sample.}
\label{gam_lumrat}
\end{figure}

Figure~\ref{gamma} shows the distribution of a power law photon index
for the entire sample. The 1$\sigma$ errors are large ($>0.3$) for
sources with small number of counts, while for the high quality data
the errors are small ($<0.1$). An average photon index for the high
S/N data is 1.78$\pm 0.24$ and a majority of sources have an index
within $\Gamma = 1.7-2$. This is higher than a typical index of 1.5
for a radio loud quasar \citep{elvis94, richards06}. As shown in
figure~\ref{gam_lumrat} there is no dependence between the photon
index and radio-to-X-ray luminosity ratio in this sample.

We studied absorption properties and searched for an additional (above
the Galactic column) absorption required by the data intrinsic to the
source, so at the fixed redshift of the source. We detected the
intrinsic absorption columns $\sim 10^{21}$~cm$^{-2}$ for three
quasars PKS~1127-145, B2~0738+313, Q1416+067 and $\sim
10^{22}$~cm$^{-2}$ for one galaxy, B2~1345+125. We note that two
quasars are known to have Damped Lyman Alpha (DLA) absorption systems
(PKS~1127-145, B2~0738+313, \cite{Bechtold01, siem2003}) and one
has a broad absorption line system (BAL) (Q1416+067
\cite{Bechtold02}). In general the upper limits for the intrinsic
absorption $< {\rm few} \times 10^{21}$~cm$^{-2}$ indicate low columns in
majority of quasars.

We do not detect any trends in absorption with photon index or source
luminosity.  A comparison with the GPS galaxies \citep{guainazzi2006}
indicates that the galaxies have larger absorption columns than the
GPS quasars. This may be due to a difference in the viewing angle
between the quasars and galaxies, so the galaxies are observed along
the direction of an obscurer, e.g. ``torus''.

\section{Discussion.}

\subsection{X-rays from Unresolved Core}

A typical radio size of the GPS/CSS source is relatively small in
comparison to the resolution of {\it Chandra} X-ray images. This means
that the X-ray emission measured in a standard source extraction
region (radius = 1.75~arcsec) contains the entire complex radio
structure of the GPS radio source (e.g. jets, hot spots, core) and we
cannot resolve the individual radio components in X-rays.  This
observational fact complicates any theoretical interpretation for the
origin of the X-ray emission. A contribution to the observed X-ray
spectrum can come not only from the central quasar power engine but
also from unresolved jets, hot spots or thermal gas heated by an
expanding radio source.

The X-ray emission could originate in an accretion flow onto a
supermassive black hole and be associated with a hot ionized medium,
e.g. corona, hot wind, jet, hot innermost flow
\citep{sobol2004a, sobol2004b}. It can also be a result of
the reflection of the primary emission off the cold (10$^5$~K) disk
flow (e.g. \cite{ross1993}). However, in radio-loud sources the jet
emission often dominates over these accretion components.
In fact the radio to gamma-rays emission is entirely dominated
by the jet emission in blazars' which are observed along the jet axes
\citep{sikora1997}. Some of the GPS sources might be indeed observed 
along the jet axis and have a significant X-ray emission due to the
relativistic jet particles.

If, for example, a relativistic jet is propagating within a strong IR
photon field the IR photons upscattered by the jet electrons can
contribute to the X-ray and gamma-ray emission (see \cite{sikora2002},
\citet{blazej}). The expected spectrum is flat in comparison to the photon
index found in our sample except for PKS~1127-145 with
$\Gamma=1.20\pm0.03$, the smallest in the sample.  The distribution of
a power law photon index for the sample is plotted in
Fig.~\ref{gamma}. A majority of the sources have $1.8 < \, \Gamma < 2$
with the median value of $\Gamma$= 1.84$\pm0.24$ for the quasars in
the sample. This $\Gamma$ is higher than the values of
1.57$\pm$0.08 \cite{ bechtold94} or 1.55$\pm0.17$ \cite{belsole2006}
observed for the other radio loud quasars
and similar to the value of 2.03$\pm0.31$ observed in radio quiet
quasars
\cite{kelly2007} where the X-ray emission is not associated with a
jet.  We conclude that there is no evidence for a strong contribution
of a parsec scale jet to the X-ray spectrum in 13/14 objects in our
sample of GPS/CSS quasars. This needs to be confirm with observations of
a larger X-ray sample of GPS/CSS sources.

\subsubsection{Expanding Radio Source}

Can a powerful expanding radio source, (unresolved by {\it Chandra}),
contribute to the X-ray spectrum?  The VLBI observations indicate
relativistic motions associated with the expanding jet components
\footnote{http://www.physics.purdue.edu/astro/MOJAVE/}. Outside a
parsec-scale region the GPS radio jets show knots and hot spots
emission. Such features indicate sites of shocks, interactions and
particle acceleration and in principle lead to the X-ray emission
through the synchrotron, inverse Compton processes or thermal emission
of hot, shocked interstellar medium.  Heinz, Reynolds and Begelman
(1998) considered the evolution of radio-source expansion within host
galaxies.  They simulated interactions between a growing radio source
and the interstellar and intergalactic medium. For the highly
supersonic expansion of the young source a shock forms around the
expanding source and it heats up the medium to X-ray temperatures. As
a result a ``cocoon'' of hot medium surrounds the radio source.
Depending on the density of the medium and the strength of the shock a
source of the size of 16~kpc can emit $\sim10^{45}$~erg~sec$^{-1}$ in
the {\it Chandra} band. Such an X-ray luminosity is of the order of
the luminosity observed for the sources in our sample
(Table~3). Recently \cite{stawarz2007} (see also \cite{perucho2002})
modeled the spectra of GPS galaxies with the emission from expanding
radio lobe and applied the model to a sample of GPS galaxies. Such
emission will be featureless.  Spectral lines would be present if the
emission originates from a hot thermal plasma and, depending on
metallicity, we would expect to detect the emission lines due to
metals, in particular oxygen and iron lines.

Thermal X-ray emission due to the shock heated plasma can be easily
confused with the emission from the accretion flow. Reflection off
cold/warm matter in the accretion disks can be present in some
sources.  A characteristic Fe K-$\alpha$ fluorescent emission line is
usually associated with the reflection component and indicates that
the X-ray emission originates in the accretion flow.  Of the 8 sources
in our sample with a good signal to noise data we detected Fe-line
emission in two quasars, Q0740+380 and Q1328+254. In both sources the
equivalent width of the emission line is relatively small, $<$0.4~keV
(90\%). The energy of the detected line, E$_{rest}$=6.40$\pm 0.06$~keV
in both cases, indicates that it is not coming from the ionized,
thermal material, but it may indicate a reflection component becoming
important in these sources.

If the intrinsic absorption is high the reflection component can be
detected at higher energies (E$_{rest}=\sim 4-10$~keV) and can provide
information on the intrinsic source luminosity. GPS galaxies are
likely to be highly absorbed \citep{guainazzi2006, vink2006}. We find
an equivalent Hydrogen column density of $N_H > 10^{22}$cm$^{-2}$ in
two galaxies
\citep{gps2003}. In contrast there is no significant absorption 
present in the observed by {\it Chandra} GPS quasars. This result is
in agreement with the other X-ray studies of radio sources where the
higher absorption has been detected in the galaxies than the quasars
\citep{belsole2006}, however, the absorption columns in GPS galaxies 
are not higher than the columns observed in other galaxies
\citep{guainazzi2006, vink2006}).

We note that we detected the absorption columns in three quasars (see
Table 2) and two are known to have intervening damped Lyman-$\alpha$
absorbers on the line of sight, while the third one an associated
absorber with metal system detectable in the optical spectrum. The
current data do not allow us to constrain the redshifts of the X-ray
absorbers, so there is no confirmation on whether the detected
absorption is due to intervening DLA systems, or intrinsic to the
quasar. The full description of detectability of the X-ray absorption
due to DLA is given in Bechtold et al (in preparation).

\subsubsection{Spectral Energy Distributions}

\begin{figure}
\includegraphics[width=8cm, angle=90]{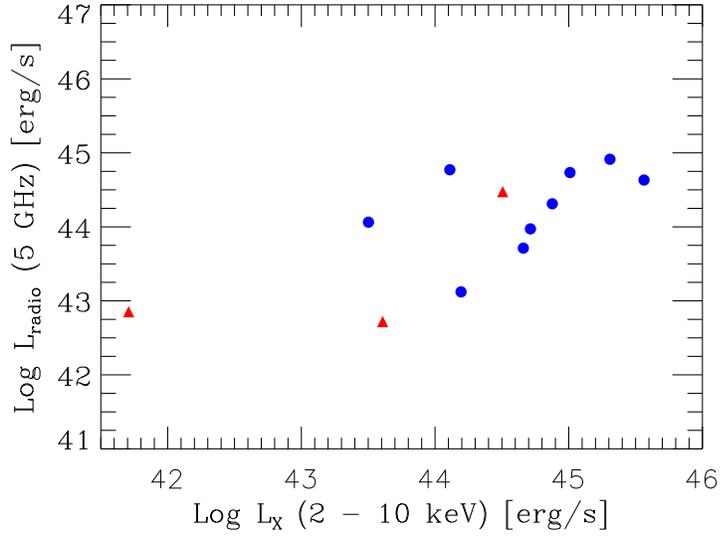}
\caption{X-ray luminosity in 2-10~keV energy range  vs. radio luminosity 
at 5~GHz.  The galaxies are marked by red triangles.}
\label{lum_x_rad}
\end{figure}

\begin{figure}
\includegraphics[width=8cm, angle=90]{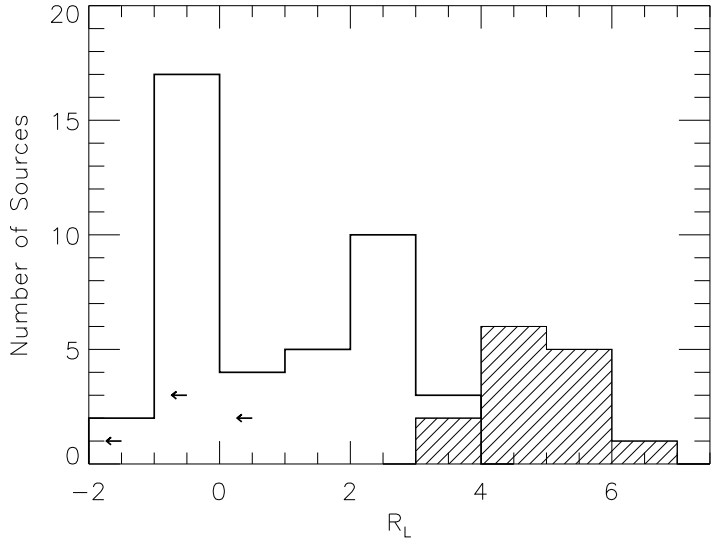}

\caption{Histogram of radio loudness. A shaded region indicate 
a parameter space for the GPS/CSS sources in our sample. For
comparison the quasars from \citet{elvis94} are shown with the thick
line with a smaller radio loudness parameter than the GPS/CSS
sample. The upper limits for the sources in \citet{elvis94} are
indicated by arrows.}
\label{radio-loudness}
\end{figure}

The optical-UV emission of the GPS/CSS sources is typical of broad
line quasars. Both broad emission lines and a big blue bump are
present in all cases and there is no signature of a jet synchrotron
emission in the optical-UV band. However, GPS sources are strong radio
emitters.  We compiled the spectral energy distribution (SED) for our
sample using the existing broad-band data available in NED.
Figure~\ref{radio-loudness} shows the radio loudness ($\rm Log
(F_{5GHz}/F_B)$) of the GPS/CSS sources in our sample in comparison to
the other radio loud quasars in \citet{elvis94}. The GPS/CSS quasars
are at the higher end of the radio loudness distribution with most of
the sources at $R_L>4$. Note that this trend is also true when we
compare GPS/CSS sources to the sample of radio loud sources presented
by \cite{sikora2007} where the calculated radio luminosity included
the entire radio source, e.g. the core, lobes and jets.
Figures~\ref{gam_lumrat} and
\ref{lum_x_rad} show the X-ray luminosity in comparison to the radio
luminosity at 5~GHz. There is no correlations between radio and X-ray
properties visible in these two figures.

\begin{figure}
\includegraphics[width=2in,angle=90]{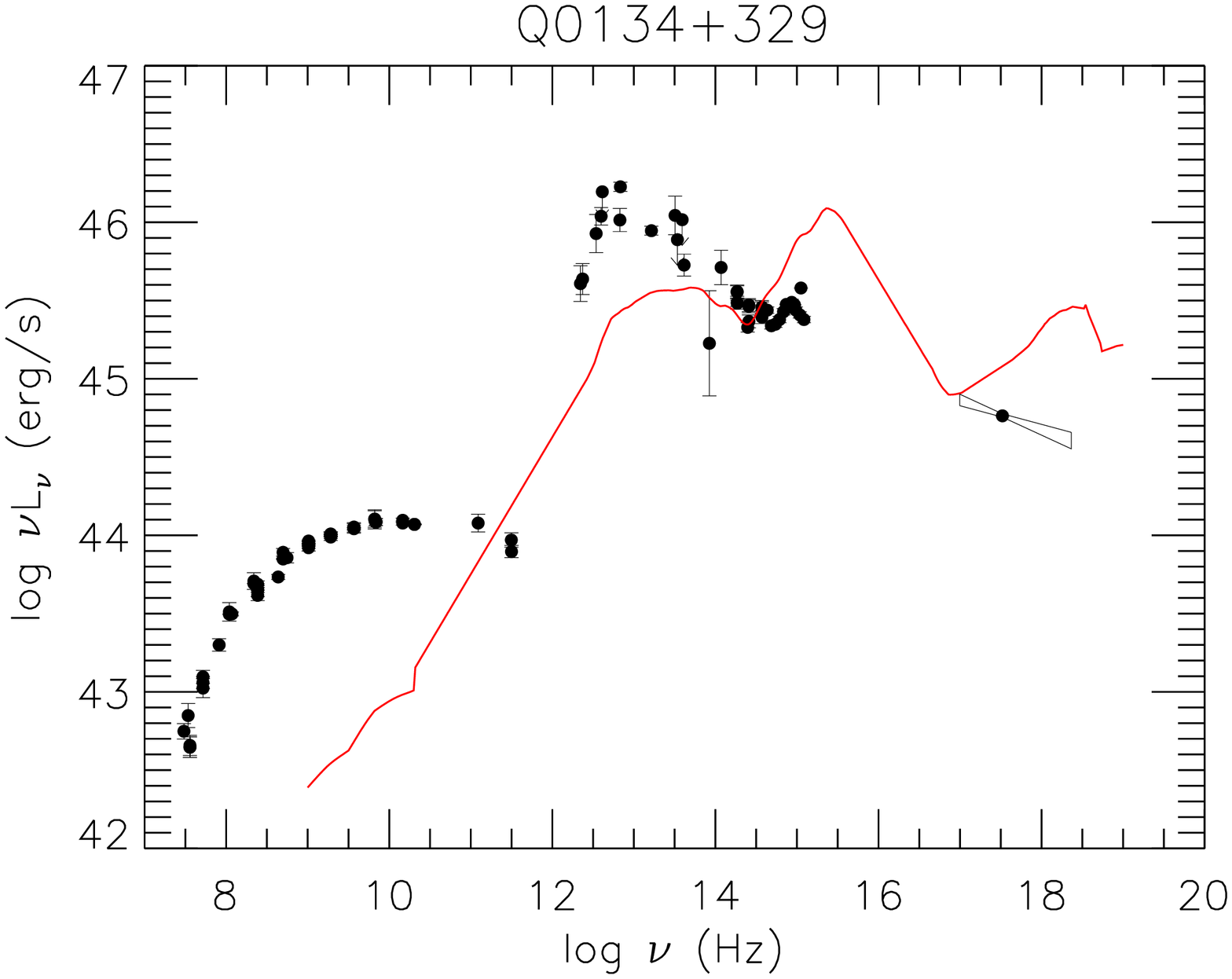}
\hfill
\includegraphics[width=2in,angle=90]{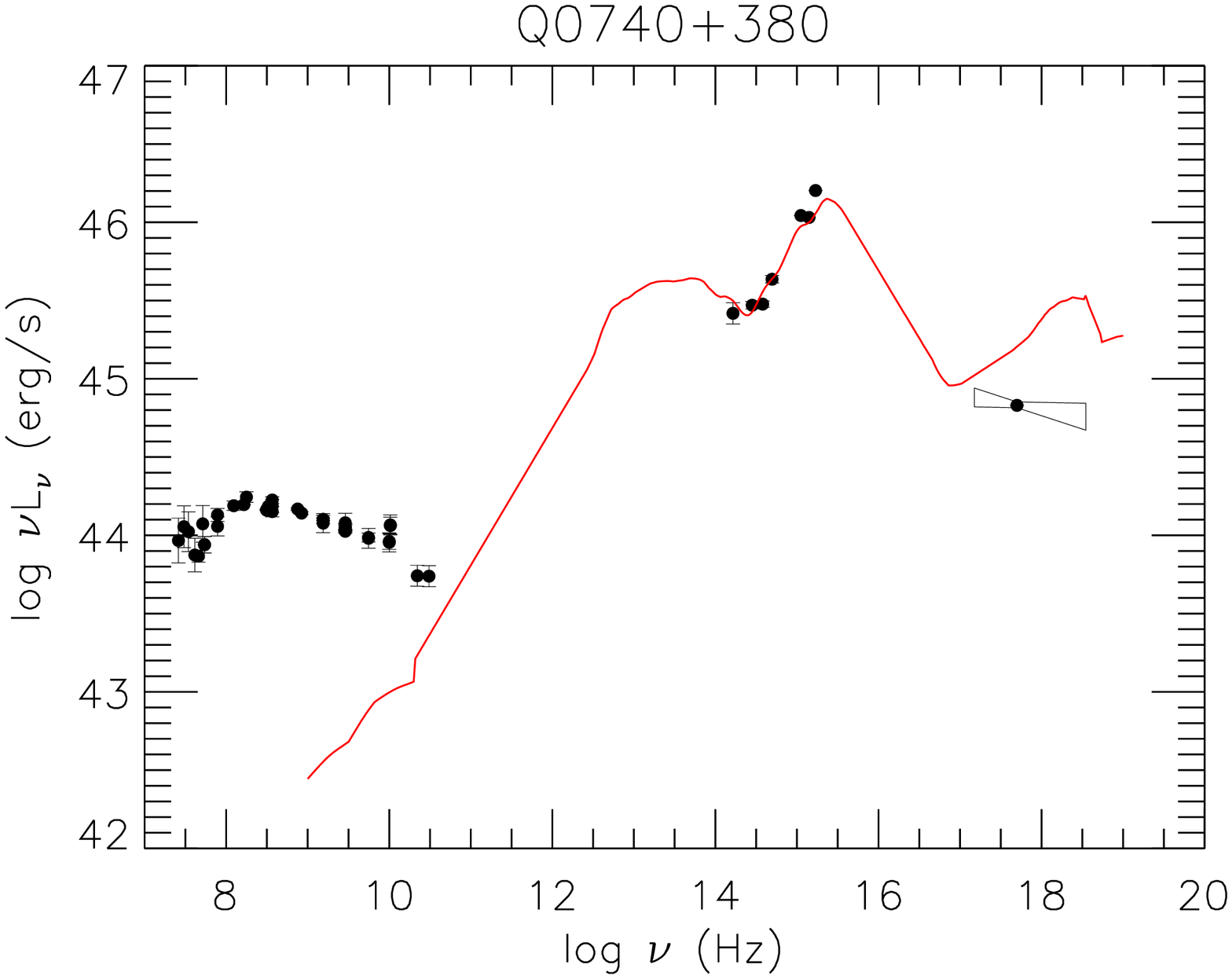}
\hfill
\includegraphics[width=2in,angle=90]{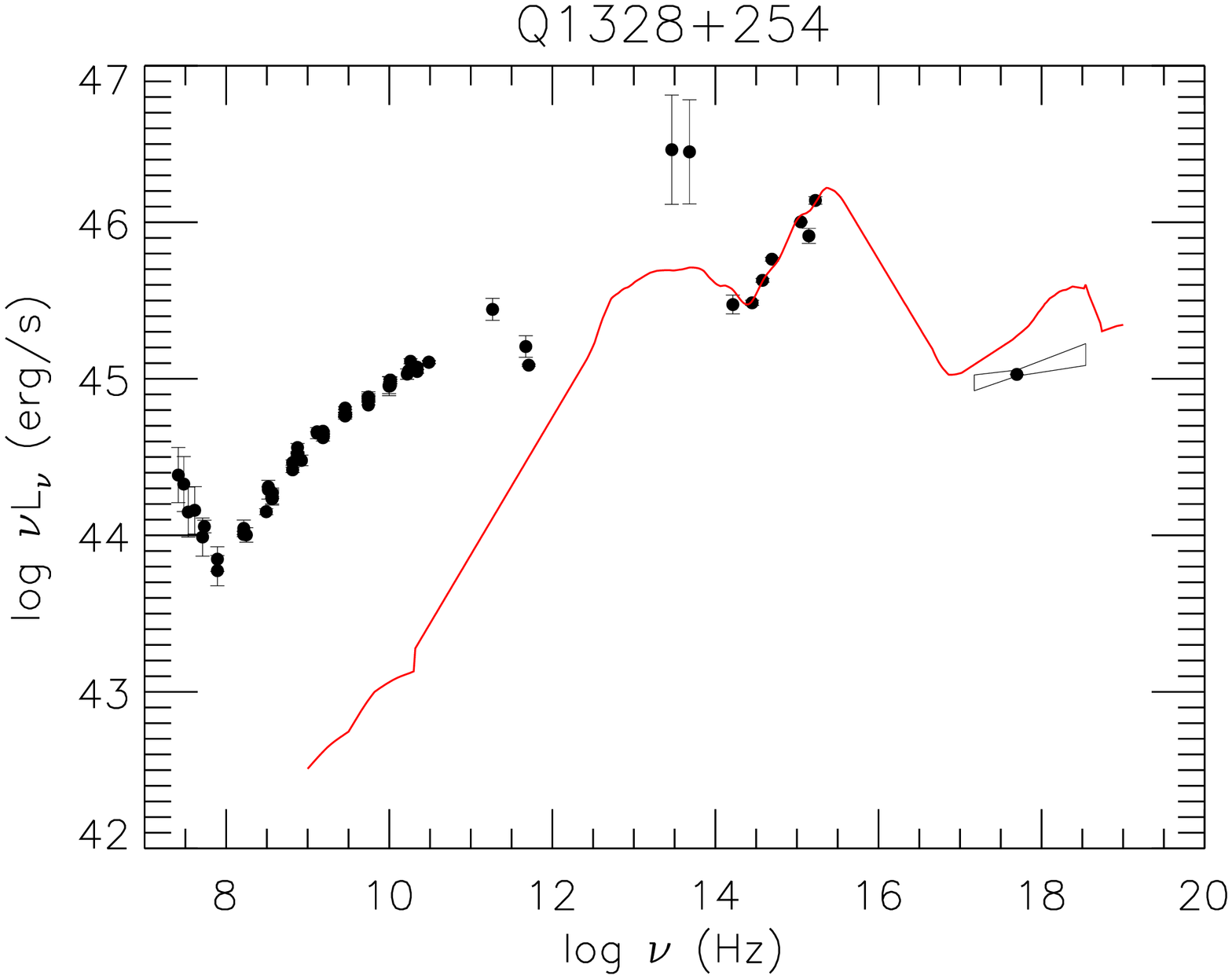}
\hfill
\includegraphics[width=2in,angle=90]{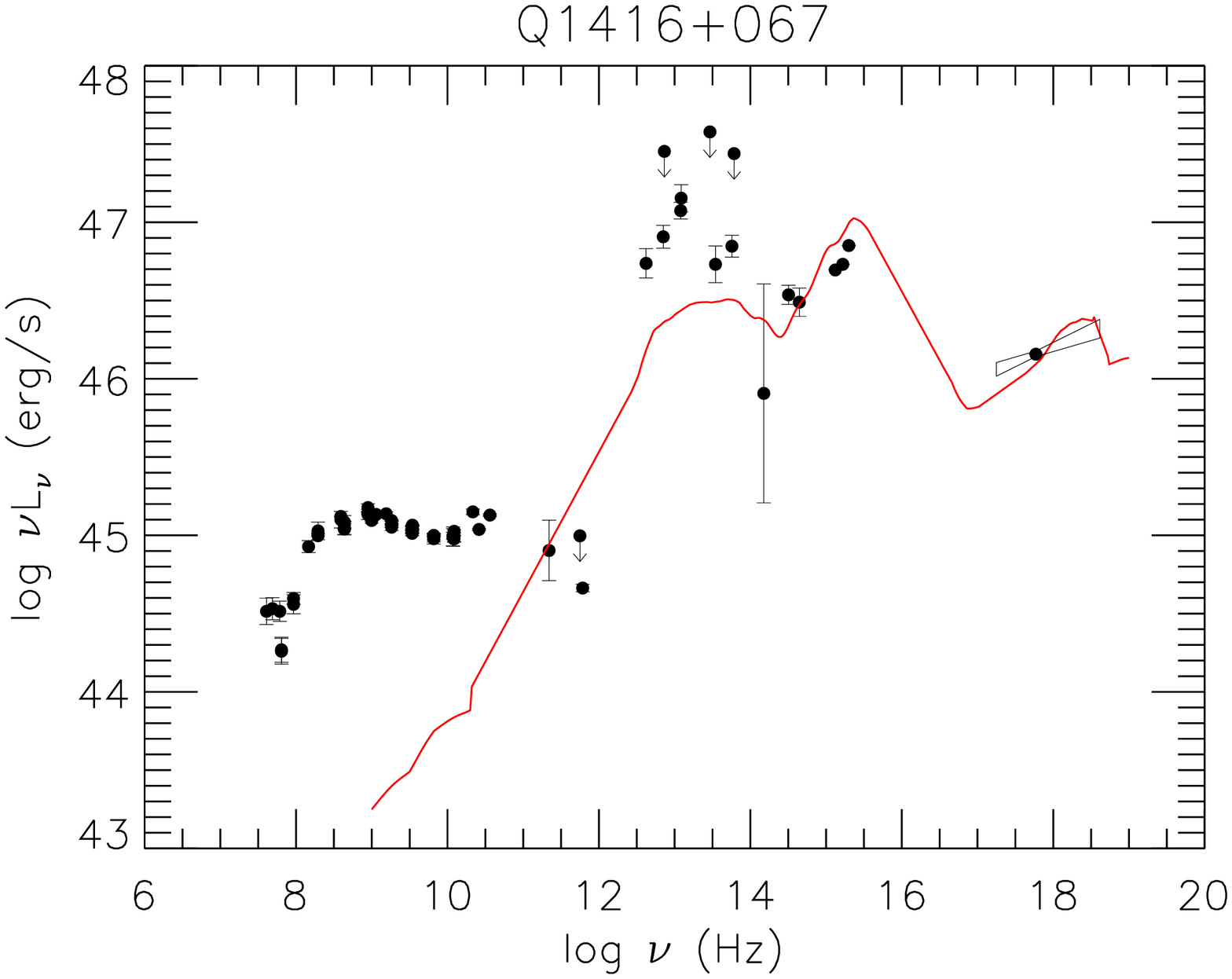}
\hfill
\caption{Examples of SEDs, where there is enough photometric data available in 
NED. The {\it Chandra} data are plotted with the 1$\sigma$ bow-tie
regions. The solid line represents the average radio-loud quasar SED
from \citet{elvis94} normalized at log$\nu$=14.5 minimum point. A
strong radio emission of these sources exceeds an average radio values
for radio-loud quasars, while the optical-UV big blue bump is typical
for the quasars.}
\label{sed}
\end{figure}

Plots in Figure~\ref{sed} shows the four quasars for which we could
build a broadband SED. The big blue bump is prominent in all
sources. For comparison we plot the radio loud SED compiled by
\citet{elvis94} normalized at 1 micron. 
The strong radio emission exceeding an average radio-loud quasar's SED
is clearly visible indicating a much stronger radio/optical power in
GPS sources, (by a factor of $\sim 30$) than in normal radio-loud
quasar. Interestingly the X-ray luminosity shown in the SEDs is
similar or even lower than that for the radio loud quasars. This might
indicate that the contribution from the GPS radio components to the
X-ray spectrum is small. We calculated the $\alpha_{ox}$ parameter for
all the sources in the sample. The values are in Table~3. The median
for the sample is 1.53$\pm0.24$. This is in agreement with
$\alpha_{ox} = 1.49\pm0.19$ values for radio quiet quasars \citep{kelly2007,
kelly2008, sobol2008} and suggests that the X-ray emission for the
sources in our sample is most likely related to the accretion process
as in radio quiet quasars. We note that PKS~1127-145 maybe an
exception because it has both a hard X-ray spectrum ($\Gamma = 1.2\pm
0.03$) and a small $\alpha_{ox}$ equal to 1.29.  Therefore X-ray
spectral analysis of a photon index and other spectral features for a
larger sample are needed to firmly established such conclusions.

\subsection{Large Scale X-ray Emission}

Three types of large scale X-ray morphology associated with
the GPS source: (1) an X-ray jet; (2) a diffuse X-ray emission
surrounding the source; and (3) a secondary source within 10-20
arcsec.

\subsubsection{Jets}

In general many detections of a large scale radio emission were not
claimed significant before the corresponding X-ray detections were
found by {\it Chandra}. High dynamic range observations are required
for detecting faint structures in the vicinity of a bright point
source, and the experiments need to be designed for specific purpose
of detecting such emission. 

Large scale X-ray jets were discovered in two sources in our sample:
PKS1127-145 \citep{siem2002,siemi2007} and B2~0738+313 \citep{siem2003}.  
A re-analysis of the radio data confirmed the presence of the
radio jets. Both jets have similar morphologies in X-ray and radio.
As in the other large scale X-ray jets \citep{jets2006} the question of the
primary mechanism responsible for the X-ray emission has not been
resolved.  The X-ray emission is modeled either as the synchrotron
emission from several populations of relativistic electrons or as a
result of the inverse Compton scattering of the Cosmic Microwave
Background (CMB) photons off relativistic electrons within the jet (see
\citet{jets2006} for the most recent review of the X-ray
jets). There is no confirming evidence favoring either of the two
models.

A large scale jet emission gives a possible evidence for the source
non-GPS status (see \cite{stan2005} for discussion on the
classification of the GPS sources). The blazar-like emission can be
responsible for the overall emission and the analysis of a 'young' GPS
class is confusing. On the other hand there should be some young
sources within the blazar class and the question is how to find them.

\subsubsection{Diffuse X-ray Emission}

Why do we expect any diffuse X-ray emission to be associated with the
GPS/CSS? There have been several possibilities considered including:
(1) a relic of the past activity, (2) a confining medium, (3)
signatures of the interactions between the jet and interstellar
medium, (4) a remnant of the past merger and (5) an X-ray cluster. 

The typical size of the GPS/CSS source is comparable to the size of
the host galaxy. However, a small fraction of the GPS/CSS sources
exhibits a large scale radio emission, which has been associated with
the past source activity \citep{baum90,kunert2005, marecki2006}. In a
few cases a young GPS source is growing within a large scale double
radio source
\citep{marecki2006}.  Do we expect any extended X-ray emission to be
present in those sources?  The old relic can produce X-rays by
upscattering CMB photons on an old population of electrons in the old
radio structures, or if there is still remaining jet activity
supplying the energy to the outer structures.  A large scale jet in
B2~0738+313 could represent such old structure \citep{siem2003}. A
faint radio lobe emission has been detected on both sides of the core,
while the X-ray jet is propagating towards one of the two lobes.
However, we do not detect any X-rays associated with the radio lobes
in this source.

Powerful radio-loud quasars are often found in rich clusters of
galaxies \citep{yee1987}.  However, there have been no systematic
studies of the X-ray environment of GPS/CSS sources. In our sample we
detect an extended diffuse X-ray emission associated with the X-ray
cluster of 3C~186, z=1.063 quasar and an extended emission in GPS
galaxy, PKS~1345+125. The properties of the 3C~186 cluster were
discussed in \citet{siem2005}. One important conclusion from the
studies of 3C~186 is that the hot cluster gas is not able to confine
the expanding radio source and that the jet moves out of the host
galaxy with no significant energy loss.

The diffuse X-ray emission in PKS~B1345+125 may originate as a thermal
emission associated with the galaxy halo.  The size of the extended
X-rays is of order $\sim 20$~kpc and it agrees with the size of the
extended emission line region of $< 20$~kpc studied in optical
\citep{holt,axon}.  \cite{holt} identified three kinematically distinct 
emission line regions in this source. They associated the narrow line
component that was the most extended one with the quiescent ISM in the
galaxy halo. The X-ray emission is elongated
towards the South-West similarly to the optical emission and it also
agrees with the VLBI jet axis \citep{stan2001}, suggesting that it is
somehow related to the expanding GPS source. More detailed discussion
of this interesting source and the signatures of the radio source
interactions with the ISM is presented in Guainazzi et al (in
preparation).

In the models for GPS/CSS source confinement the required density to
significantly slow down the jet is relatively high
(\cite{deyoung93}). Such a dense medium could be detected in X-rays
through the intrinsic absorption (or emission). While X-ray
observations of GPS galaxies suggest that they are highly obscured
\citep{guainazzi2006,vink2006} the GPS quasars in this sample
do not show significant X-ray absorption.

\subsubsection{Binary}

Some GPS sources show signatures of mergers, although at
the same level as other active and non-active galaxies
\citep{devries2000}. A radio source in the vicinty of 
PKS~0941-080 (z=0.228) was identified by \citet{stan2005} in their VLA
image at about 20~arcsec ($\sim 50$~kpc) to the West of the double
nucleus galaxy \citep{devrie}. There is no optical counterpart to this
radio source.  In 5~ksec {\it Chandra} observation of PKS~0941-080 we
detected both radio sources, but only 4 counts were detected at the
location of the second radio source. This observation was not deep
enough to study an X-ray environment of this possibly interacting
system.  The GPS galaxy PKS~0941-080 is the faintest $f_X(0.5-2~keV) =
5 \times 10^{-15}$erg~cm$^{-2}$~s$^{-1}$ and the least X-ray luminous
between all GPS galaxies detected so far in X-rays ($L_X(0.5-2~keV = 6
\times 10^{41}$erg~s$^{-1}$) \citep{guainazzi2006, vink2006}.
No other source in the sample has a second component detected in
X-rays.

\subsection{Jets and Accretion Power: the Evolution of Radio Sources}

In terms of the evolution of radio sources, the GPS source might
represent an early stage of the quasar activity. A process responsible
for a triggering of the quasar activity remains one of the key open
questions. A central supermassive black hole requires a large amount
of fuel to power a quasar.  Recent simulations of hierarchical
structure formation suggest that the quasar activity is a direct
consequence of galaxy interactions resulting in a rapid fuel supply to
the central black hole.  On the other hand some ``feedback'' or other
intermittency mechanism (see, e.g. Janiuk et al 2004, Siemiginowska et
al 1996), for accretion flow instabilities) is required to induce the
intermittent source activity observed in nearby clusters of galaxies
\citep{clusters2007}.

If GPS/CSS sources are representative of an initial activity stage
then one would expect to detect some signatures of the increased fuel
supply into the central regions of their host galaxy. The estimated
age of the smallest GPS galaxies from the expansion velocity of the
radio components indicates ages below 10$^3$~yrs \citep{polatidis2003}
while the synchrotron ages for larger samples show usually age
$<10^5$~years \citep{murgia}.

\cite{sikora2007} studied a relationship between the source 
radio power and the accretion power (defined as Eddington luminosity)
for a large sample of AGN. They compare a total radio power of the
source (nucleus and the large scale radio emission) to the accretion
power of the nucleus and show that the radio luminosity is relatively
constant for large optical luminosities, but decreases for the low
optical luminosities. The sources in our sample are extremely radio
luminous and radio-loud Figure~\ref{radio-loudness} shows that the
radio-loudness for the sources in our sample exceeds the
radio-loudness of the radio-loud quasars in \citet{elvis94}. Their
optical to X-ray luminosities expressed as $\alpha_{ox}= 1.53\pm0.24$
are instead within the range values of radio quiet quasars
1.49$\pm0.19$ \citep{kelly2007} suggesting that the X-ray emission
processes are most likely related to the accretion process. The high
radio-loudness may be related to the higher efficiency of radiating
the jet power in compact radio sources.

\section{Summary}

We discussed {\it Chandra} X-ray observations of a sample of GPS/CSS
sources. We detected all sources and study their X-ray spectral and
spatial properties. GPS quasars are not absorbed in contrast to GPS
galaxies that show high X-ray column densities. The median X-ray
photon index 1.84$\pm0.24$ is steeper than the one observed in radio
loud quasars, while the optical to X-ray luminosity ratios
$\alpha_{ox}= 1.53\pm0.24$ are typical of radio quiet quasars. We may
conclude that the X-ray emission in these sources is most likely
related to the accretion power as in radio quiet quasars and not to
the relativistic jet, except for PKS~1127-145.

\acknowledgments

We thank the referee for insigthful comments.  AS thanks Diana
Worrall, Matteo Guainazzi, Marek Sikora, Lukasz Stawarz and Carlo
Stanghellini for discussions and comments.  This research is funded in
part by NASA contract NAS8-39073. Partial support for this work was
provided by the National Aeronautics and Space Administration through
Chandra Awards Number GO-01164X, GO2-3148A, GO5-6113X issued by the
Chandra X-Ray Observatory Center, which is operated by the Smithsonian
Astrophysical Observatory for and on behalf of NASA under contract
NAS8-39073. This work was supported in part by NASA grants
GO-09820.01-A and NAS8-39073

\clearpage

\begin{table}
\caption{GPS/CSS Chandra Sample}
\medskip
\begin{scriptsize}
\label{tab1}
\begin{tabular}{llclcllccclcl}
\hline
\hline\noalign{\smallskip}
Name & & Type\tablenotemark{a} & redshift & N$_{H}$\tablenotemark{b} & Radio\tablenotemark{c} & Radio & Exposure &
Chandra & OBSID & Tot cts & Net cts\\ & & & & 10$^{20}$ cm$^{-2}$ & Size & Morph & ksec & Obs Date \\
\hline\noalign{\smallskip}
0134+329 & 3C48 & Q/CSS & 0.367 & 4.54 & 0.5  & CJ$^5$ & 9.2 & 2002-03-06 & 3097 & 6726 & 5318.8 \\
0615+82  &      & Q/GPS* & 0.71  & 5.3  & 0.5 & IR$^2$ & 47.3 & 2001-10-18 & 1602 & 2395 & 2178.0 \\
0738+32  &      & Q/GPS & 0.63  & 4.18 & 0.01 & CJ$^1$ &  27.6 & 2000-10-10 & 377  & 3675 & 3431.2  \\ 
0740+380 & 3C186 &Q/CSS & 1.063 & 5.64 & 2.2  & DLJ$^7$ & 34.4 & 2002-05-16 & 3098 & 1830 & 1702.2 \\
0941-080 &       &G/GPS & 0.228 & 3.67 & 0.05 & CSO$^1$& 5.35  & 2002-03-26 & 3099 & 9 & 8.8 \\
1127-145 &      & Q/GPS & 1.18  & 3.12 & 0.003 & CJ$^1$ & 27.3 & 2000-05-28 &  866 & 14,972 & 14,972.6 \\ %
1143-245 &      & Q/GPS & 1.95  & 5.22 & 0.006 & CJ$^1$  & 4.95 & 2002-03-08 & 3100 & 222 & 219.7 \\
1245-197 &      & Q/GPS & 1.28  & 4.72 & 0.5  & CSO$^6$ & 5.1  & 2001-12-23 & 3101 & 44 & 43.8  \\
1250+568 & 3C277.1 & Q/CSS & 0.32  & 1.22 & 1.67& DLJ$^5$ & 14.0 & 2002-10-27 & 3102 & 2337 & 2277.1 \\
1328+254 & 3C287   & Q/CSS & 1.055 & 1.08 & 0.048& CJ$^2$&36.2& 2002-01-06 & 3103 & 3509 & 3415.6 \\
1345+12  & 4C12.50 &G/GPS & 0.122 & 1.9   & 0.08 & CSO$^1$ &25.3& 2000-02-24 & 836 & 1347 & 1335.2 \\ 
1416+067 & 3C298   &Q/CSS & 1.439 & 2.5   & 1.49 & DLJ$^5$ &  17.9 & 2002-03-01 & 3104 & 10,183 & 9522.1 \\
1458+718 &  3C309.1 & Q/CSS & 0.905 & 2.33 & 2.11& CSO$^8$ & 16.95 & 2002-01-28 & 3105 & 5434 & 5104.6 \\
1815+614 &         & Q/GPS & 0.601 & 3.9   & & CSO$^3$ &  4.9   & 2002-09-25 & 3056 & 164 & 142.0  \\
1829+29  & 4C29.56 & Q/CSS & 0.842 & 11.16 & 3.1 &  CSO$^4$ & 5.3 & 2002-10-08 & 3106 & 19 & 18.9 \\
2128+048 &         & G/GPS & 0.99  & 5.23  & 0.03 &CSO$^1$ &5.7& 2002-10-11 & 3107 & 92 & 90.7 \\

\hline
\hline\noalign{\smallskip}
\end{tabular}
\end{scriptsize}
\tablenotetext{a}{G-Galaxy, Q-Quasar, GPS or CSS radio classification based on O'Dea (1998).}
\tablenotetext{b}{equivalent Hydrogen column in the Milky Way from COLDEN \citep{stark1992}.}
\tablenotetext{c}{radio size in arcsec from O'Dea (1998);}
\tablenotetext{d}{radio morphology defined as: CJ- core-jet, DLJ- double lobes -
jet, CSO-compact symmetric object,IR - irregular;} 
\tablerefs{
$^1$ Stanghellini et al (2005); $^2$ \cite{kallerman98};  
\cite{edwardstingay}; $^3$ Taylor et al (1994); $^4$ \cite{dallacasa};
$^5$ \cite{Akujor}; $^6$ \cite{taylor2003} not definite classification for this source. 
$^7$ \cite{cawthorne}; $^8$ \cite{Pearson98}.}
\end{table}

\begin{table}
\scriptsize
\begin{center}
\caption{Absorbed Power Law Model Fits.}
\medskip
\label{tab2}
\begin{tabular}{lcccccl}
\hline
\hline\noalign{\smallskip}
 & gal & abs & abs & $\Gamma$ & Norm & $\chi^{2}$\\
 & nH  & nH & Redshift &  & [1 keV]  & (DOF)\\
 & $10^{20}$ & $10^{20}$ &  & & 10$^{-5}$ph/cm$^{2}$/s \\
\hline

Q0615+820 & 5.3 & $<$4.69 & 0.770 & 1.73$_{-0.06}^{+0.06}$ &
 5.63$_{-0.22}^{+0.23}$ & 90.6(101)\\ 

B2 0738+313 & 4.2 & 8.3$_{-6.1}^{+6.4}$ & 0.630 &
1.56$_{-0.08}^{+0.08}$ & 13.80$_{-0.98}^{+1.05}$ &  172.4(143)\\ 

Q0740+380 & 5.6 & $<$8.44 & 1.063 &
2.09$_{-0.08}^{+0.08}$ & 6.92$_{-0.28}^{+0.31}$ & 73.3(78)\\ 

Q1127-145  & 4.1 & 17.8$_{-6.6}^{+6.5}$ & 1.180 &
1.20$_{-0.03}^{+0.03}$ &  52.69$_{-1.75}^{+1.80}$ & 346.2(310)\\

Q1143-245 & 5.2 & $<$76.88 & 1.950 &
1.62$_{-0.22}^{+0.26}$ &  5.09$_{-0.78}^{+1.00}$  & 11.1(9)\\

Q1250+568& 1.2 & $<$2.25 & 0.320 &
1.85$_{-0.07}^{+0.07}$ &  19.27$_{-0.71}^{+0.75}$ & 112.7(98) \\

Q1328+254 & 1.1 & $<$15.75 & 1.055 &
1.86$_{-0.05}^{+0.07}$ & 11.12$_{-0.33}^{+0.66}$ &  111.5(124)\\

PKS B1345+125$^G$  & 1.9 & 254.3$_{-58.0}^{+63.6}$ &  0.122 &
1.10$_{-0.28}^{+0.29}$ & 12.11$_{-4.16}^{+6.57}$ & 107.8(71)\\

Q1458+718 & 2.3 & $<$3.19 & 0.905 &
1.57$_{-0.04}^{+0.04}$ &  33.45$_{-0.88}^{+0.90}$ & 165.2(168)\\

\hline
\multicolumn{7}{c}{Targets with low counts (N$<200$) fit with N$_H$ fixed at the Galactic column } \\
\hline

Q0941-080$^G$ & 3.7 & & & 2.62$_{-1.03}^{+1.29}$ &
0.25$_{-0.11}^{+0.16}$ & 49.0(442)\\

Q1245-197  & 4.7 & & & 1.96$_{-0.43}^{+0.45}$ & 1.17$_{-0.29}^{+0.34}$ &
 132.5(442)\\

Q1815+614 & 3.9 & & & 1.70$_{-0.31}^{+0.33}$ & 3.42$_{-0.62}^{+0.61}$
& 5.5(6)  \\ 

Q1829+290 & 11.2 & & &1.82$_{-0.66}^{+0.72}$ & 0.54$_{-0.21}^{+0.28}$
& 98.2(442) \\ 

PKS B2128+048$^G$ & 5.2 & & & 1.28$_{-0.41}^{+0.42}$ &
1.58$_{-0.38}^{+0.37}$ & 6.1(3)\\

\hline
\multicolumn{7}{c}{Targets affected by pile-up.}\\
\hline

Q0134+329 & 4.5 & $<$0.75 & 0.367 & 2.19$_{-0.05}^{+0.05}$ & 79.89$_{-1.96}^{+1.92}$ & 212.4(143)\\ 

	& & & & 2.52$_{-0.09}^{+0.07}$& 91.0$_{-3.2}^{+0.1}$ & 180.9(141)\\



Q1416+067 & 2.2 & 27.1$_{-9.7}^{+10.0}$ & 1.439 &
1.85$_{-0.05}^{+0.05}$ &  70.04$_{-2.82}^{+2.93}$ & 226.7(203) \\

	& \\
\hline
\hline
\end{tabular}

$^G$ Galaxy; Errors are 90$\%$ for one significant parameter.
The upper limits are 3$\sigma$.
\end{center}
\end{table}




\begin{table}

\scriptsize
\label{tab4}
\caption{Flux and Luminosity for GPS/CSS Chandra Sample}
\begin{tabular}{lccccccccccl}
\hline
\hline\noalign{\smallskip}
 & Flux & L$_{x}$ & Flux & L$_{x}$ & L$_{B}$ & L$_{R}$ & R$_{X}$$^{1}$ &
R$_{L}$$^{2}$ & $\alpha_{ox}$$^{3}$ & $\alpha_{ro}$$^{4}$ & \\ 
 & [0.5-2keV] & [0.5-2keV] & [2-10keV] & [2-10keV] & & [5 GHz] & & & & \\
 & $10^{-13}$erg/cm$^{2}$/s & $10^{44}$erg/s &
$10^{-13}$erg/cm$^{2}$/s & $10^{44}$erg/s &  $10^{44}$erg/s  &$10^{44}$erg/s \\ 
\hline
\hline

Q0134+329 & 15.22 & 2.89 & 15.58 & 5.17 & 11.66 & 0.94 & 7.36 &  4.28 & 1.54 & 0.63 
 \\ 

Q0615+820$^{5}$ & 1.07 & 0.70 & 2.24 & 3.46  & 2.98 & 0.87 & 7.64  & 5.05  & 1.35 &
0.55  \\ 

B2 0738+313 & 3.12 & 1.46 & 7.37 & 7.53 & 1.65 & 2.05 & 7.76 & 5.61  & 1.72 & 0.60 \\ 

Q0740+380 & 1.28 & 1.39 & 1.55 & 4.57  & 13.37 & 0.52 & 7.14 & 4.29  & 1.74 & 0.50 \\ 

Q0941-080 & 0.05 & 0.006 & 0.03 & 0.005 & 0.004 &0.07 & 9.31 &  6.53  & 2.11 & 0.72 \\ 

Q1127-145 & 12.22 & 15.74 & 49.38 & 179.41 & 80.82 & 8.15 & 7.10 & 4.74 & 1.29 & 0.65 \\

Q1143-245 & 1.03 & 2.91 & 2.17 & 20.33 & 7.13 & 8.20 & 7.81 & 6.04  & 1.56  & 0.68 &
\\


Q1245-197 & 0.22 & 0.33 & 0.32 & 1.29 & 0.80 & 5.91 & 8.77 & 6.65  & 1.57 & 0.88 & \\

Q1250+56  & 4.12 & 0.62 & 6.27 & 1.56  & 2.18 & 0.13 & 7.15 & 4.13 & 1.44 & 0.63 & \\

Q1328+254 & 2.39 & 2.57 & 3.51 & 10.22 & 15.70 & 5.41 &7.90 & 5.23 & 1.59 & 0.70 & \\

PKS B1345+125  & 4.04 & 0.11 & 12.18 & 0.41 & 10.08 & 0.05 & 7.53 & 3.67 & 1.52 &
0.66& \\

Q1416+067 & 15.33 & 26.96 &  23.69 & 126.23 & 98.68 & 4.81 & 6.76 & 4.32  & 1.42 &
0.57 & \\

Q1458+718 & 7.04 & 5.93 & 17.05 & 36.55 & 173.46  & 4.29 & 7.37 & 4.03 & 1.47 & 0.64 \\

Q1815+614$^{5}$ & 0.73 & 0.32 & 0.12 & 1.13 & 0.02 & 0.24 & 7.50 & 5.50 & -- & -- & \\ 

Q1829+290 & 0.09 & 0.07 & 0.18  & 0.32 & 1.16 & 1.16 & 8.73 & 6.17  & 1.73 & 0.79 \\

PKS B2128+048 & 0.32 & 0.31 &  1.25 & 3.21 & 1.23 & 2.99 & 8.37 & 6.06 & 0.91 & 1.09&\\

\hline
\hline

\multicolumn{12}{l}{
$^{1}$ $R_{X}$ = log(Flux$_{5GHz}$/Flux$_{2keV}$)}\\

\multicolumn{12}{l}{
$^{2}$ Radio loudness: log(Flux$_{5GHz}$/Flux$_{BBand}$)}\\

\multicolumn{12}{l}{
$^{3}$ $\alpha_{ox}$:
 log[Flux(2500\AA)/Flux(2keV)]/2.605, in rest frame}\\

\multicolumn{12}{l}{
$^{4}$ $\alpha_{ro}$:
 -log[Flux(2500\AA)/Flux(5GHz)]/5.38 in rest-frame.}\\
\multicolumn{11}{l}{
Calculated flux at 2500 \AA  by using V magnitude from O'Dea et
al. 1998 and from the relation: f$_\nu \sim \nu^{-\alpha}$, where
$\alpha = 0.5$}\\
\multicolumn{12}{l}{
The V magnitudes for Q0615+820 and Q1231+481 came from Veron-Cetty et
al., 2001}\\


\multicolumn{11}{l}{
$^{5}$ Radio flux values at 5GHz calculated from NED data for these
  sources; we used published values for radio flux from O'Dea et} \\

\multicolumn{11}{l}{
  al. 1998 for the other sources.}\\

\multicolumn{11}{l}{
All luminosity values are unabsorbed and in rest-frame.}\\
\end{tabular}

\end{table}


\clearpage


\begin{figure}
\includegraphics[width=8cm, angle=90]{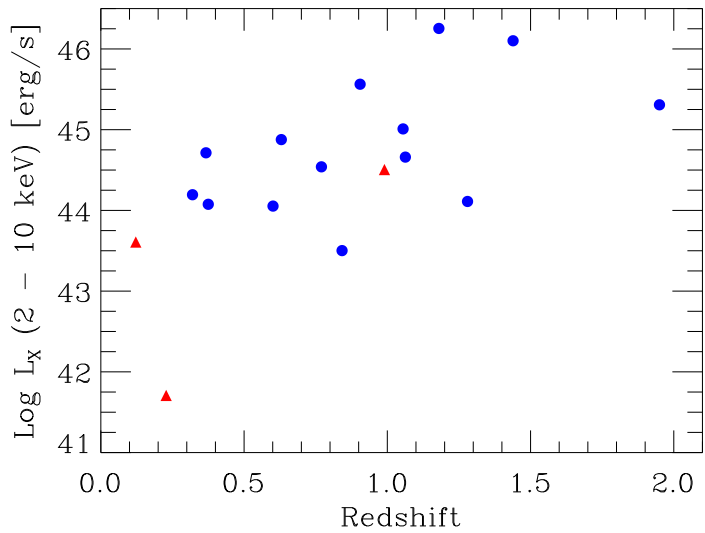}
\caption{X-ray luminosity in 2-10~keV energy range vs. redshift for the GPS/CSS sample.
The galaxies are marked with red triangles.}
\label{lumx_redshift}
\end{figure}

\end{document}